\documentclass[10pt]{iopart}
\usepackage{iopams}
\usepackage{graphicx}
\usepackage{subfigure}
\usepackage{caption}
\usepackage{epstopdf}
\usepackage{epsfig}

\usepackage{amssymb}
\usepackage{amsthm}

\begin{document}

\title[Exploring quantum phase transition in Pd$_{1-x}$Ni$_x$ nanoalloys]{Exploring quantum phase transition in Pd$_{1-x}$Ni$_x$ nanoalloys}

\author{P. Swain$^{1*}$, Suneel K. Srivastava$^2$ and Sanjeev K. Srivastava$^{1\dagger}$}
\address{$^1$ Department of Physics, Indian Institute of Technology Kharagpur, Kharagpur-721302, INDIA}
\address{$^2$ Department of Chemsitry, Indian Institute of Technology Kharagpur, Kharagpur-721302, INDIA}
\ead{$^{1*}$priya@phy.iitkgp.ernet.in,$^2$sunit@chem.iitkgp.ernet.in
and $^{1\dagger}$sanjeev@phy.iitkgp.ernet.in} \vspace{10pt}
%\begin{indented}
%\item[]
%\end{indented}

\begin{abstract}

Pd$_{1-x}$Ni$_x$ alloy system is an established ideal transition metal system possessing a composition induced paramagnetic to ferromagnetic quantum phase transition (QPT) at the critical concentration $x_c \sim$ 0.026 in bulk. A low-temperature non-Fermi liquid (NFL) behaviour around $x_c$ usually indicates the presence of quantum criticality (QC) in this system. In this work, we explore the existence of such a QPT in nanoparticles of this alloy system.
We synthesized single-phase, polydispersed and 40-50 nm mean diameter crystalline nanoparticles of Pd$_{1-x}$Ni$_x$ alloys, with $x$ near $x_c$ and beyond, by a chemical reflux method. In addition to the determination of the size, composition, phase and crystallinity of the alloys by microscopic and spectroscopic techniques, the existence of a possible QPT was explored by resistivity and DC magnetization measurements. A dip in the value of the exponent $n$ near $x_c$, and a concomitant peak in the constant $A$, of the $AT^n$ dependence of the low temperature ($T$) resistivity indicate the presence of a quantum-like phase transition in the system. The minimum value of $n$, however, remains within the Fermi liquid regime ($n >$ 2). The DC magnetization results suggest an anticipatory presence of a superparamagnetic to ferromagnetic QPT in the mean-sized nanoparticles. The observation of a possible quantum critical NFL behaviour ($n <$ 2) through resistivity is argued to be inhibited by the electron-magnon scatterings present in the smaller nanoparticles.

\end{abstract}

%\noindent{\it Keywords}: XXXXXX, YYYYYYYY, ZZZZZZZZZ \\
%\submitto{\NT}
\maketitle
\ioptwocol

\section{Introduction}

Pure Pd metal is known to be an exchange-enhanced paramagnet right on the verge of being a ferromagnet due to its high electron density of states at the Fermi energy and the resulting large Stoner enhancement factor ($\sim$ 20) \cite{Doniach67}. Isolated magnetic 3$d$ impurities like Mn, Fe and Co produce strong ferromagnetic spin polarization of the Pd $d$ electrons \cite{Oswald86, Mohn93, Swieca97}. This polarization extends over several atomic lengths around the impurity and results in giant moments. Thus, even an extremely small amount ($<$ 0.1$\%$) of these impurities produces a long-range ferromagnetic order \cite{Burke82}. When the impurity is Ni, however, it merely increases the enhanced susceptibility of Pd further, up to a relatively high impurity concentration $x_c \sim$ 2.6 $\%$. A ferromagnetic order sets in above this concentration \cite{Murani74}, the FM to PM transition temperature T$_c$ for which increases from 0 K at $x_c$ to an $(x-x_c)^{3/4}$ dependence beyond and in the vicinity of $x_c$ \cite{Nicklas99}. This ferromagnetism occurring at the relatively high Ni concentrations is known to arise from nucleations of giant moments around groups of a number of Ni atoms and the crossing of a percolation threshold \cite{Burke82} at and beyond $x_c$. This way, the Pd$_{1-x}$Ni$_x$ alloys undergo a quantum (T = 0 K) phase transition (QPT) from a PM state for $x < x_c$ to a FM state for $x > x_c$, $x_c$ being known as the quatum critical concentration (QCC) \cite{Burke82, Murani74, Nicklas99}. The occurrence of this quantum criticality (QC) in the Pd$_{1-x}$Ni$_x$ alloys has been seen experimentally at finite, though low, temperatures as deviations of temperature dependencies of macroscopic properties, like resistivity, heat capacity and magnetic susceptibility, from Fermi liquid behaviour \cite{Nicklas99}. The manifestation of QC in these alloys has also been seen even at room temperature (RT) and beyond through a microscopic experimental observation of an anomalous Gaussian-like bahaviour of the local moment of an additional isolated impurity, viz. Fe, in the alloy around the QCC \cite{Srivastava08}.

It is quite intriguing to extend the investigation of QC  in nanoparticles of Pd$_{1-x}$Ni$_x$ alloys in light of the fact that many properties of materials at nanoscale are different from their bulk counterparts due to quantum confinement effects. Pd$_{1-x}$Ni$_x$ nanoalloys are of current scientific interest because of their catalytic and magnetic properties. Hitherto existing reports on magnetism of Pd$_{1-x}$Ni$_x$ nanoalloys include (i) high Ni concentration ($\geq 20 \%$) nanoparticles/nanorods of diameter 20-30 nm which showed FM order up to RT \cite{Fanrong10, Bagaria}, exactly as reported for high Ni concentrations in bulk, and (ii) ultrafine (2-5 nm) nanoclusters of pure Pd, the surface atoms of which are bonded with some ligands \cite{Angappane08, Coronado08}. These clusters were found to be superparamagnetic with FM phase existing up to RT. Prima facie, the room temperature FM order shown by even the pure Pd particles leaves no scope for getting a 0 K non-magnetic state required for a QPT in Pd$_{1-x}$Ni$_x$ nanoparticles at any Ni concentration. However, such an investigation with single-phase, 40-50 nm Pd$_{1-x}$Ni$_x$ ($x \approx x_c$) particles without any core-shell or ligand-bond structure is still open. Although there exists another set of reports of magnetism in ultrafine naoclusters of these alloys, including the vicinity of $x_c$ \cite{Teranishi99, NunomuraPL98, NunomuraJMMM98}, the studies are focussed merely on the formation of giant moments around Ni atoms and do not address the possibility of a QPT. As far as synthesis of single-phase Pd$_{1-x}$Ni$_x$ nanoparticles in the desired composition range is concerned, there exists a report by Lee {\it et. al.} \cite{Lee2012}, wherein they performed simultaneous reduction of metal precursors of both Pd and Ni. However, the sizes were limited to $\sim$ 5 nm. Others report the synthesis of either a different (core-shell \cite{Teranishi99}, surface-layer protected \cite{NunomuraPL98, Fanrong10}, C nanotube protected \cite{Li2011}) structure, or use nuclear radiation \cite{Oh2008, Zhang2010}, which, in general, is undesirable. We, therefore, need to look for a new chemical route for the synthesis of the desired nanoalloys.

In this work, we synthesize nanoparticles of Pd$_{1-x}$Ni$_x$ alloys in the vicinity of $x_c$ and beyond, with a focus on achieving good crystallinity with single phase, using a chemical reflux method. Nanoalloys with compositions close to $x_c$ were prepared for the study of transport and magnetic properties with the objective of seeking a possible QPT. Since the $x_c$ is small, and the composition intervals in its vicinity have to be even smaller, a precise determination of the compositions is crucial. Therefore, samples with higher Ni concentrations were also prepared to verify especially the synthesized compositions ($x_S$) by finding a scaling behaviour between $x_s$ and the initial composition $x_i$ taken for the synthesis. We then verify the sizes, the phase, the crystallinity and the compositions by various microscopic and spectroscopic techniques, and then seek the possibility of a QPT by simple electrical resistivity and DC magnetization studies.

\section{Experimental Section}

The nanoalloys of Pd$_{1-x}$Ni$_x$ (0.01 $\leq x \leq$ 0.50) were prepared by a simultaneous reduction of different Pd- and Ni-ion ratios by hydrazine hydrate in the presence of surfactant diethanolamine, the reaction taking place in a conventional reflux apparatus. The following steps were followed in a typical synthesis process:  1.5 mmol of the Pd metal precursor salt Palladium(II) chloride (PdCl$_2$) was first dissolved in 24 mL of 2 M HCl in a round-bottom flask of 100 ml capacity and  subsequently stirred. This results in the formation of the Pd complex [PdCl$_4$]$^{2-}$ of Pd$^{2+}$ ions in the solution. The procedure was adopted from a work by Nguyen {\it et al.} \cite{Nguyen10} for synthesizing Pd naoparticles. Next, an appropriate amount of the Ni metal precursor salt Nickel(II) chloride hexahydrate (NiCl$_{2}$.6H$_{2}$O) was dissolved in water to yield the complex NiCl$_2$ of Ni$^{2+}$ ions in the solution. This particular reaction has been used by many including Yuan {\it et al.} \cite{Yuan2011}. The two solutions were then mixed together into one, to which 5 ml of diethanolamine (DEA) was added as a surfactant. DEA has been reported to be used for syntheses of oxide nanoparticles \cite{Qiao2009, Choi2007}; we use it to synthesize metallic alloy nanoparticles for the first time. Further, 5 ml of hydrazine hydrate was added as the common reducing agent to this solution. Adoption of this reducing agent was inspired by its use in synthesizing metallic super paramagnetic Ni nano particles by Lanje {\it et al.} \cite{Lanje2010}. In the last reaction step, 40 ml distilled water was added to this and the resulting solution was refluxed for 24 h at 110 $^{\circ}$C in an oil bath. The black coloured product formed in the process, i.e., the alloy, was then filtered, washed with distilled water, and dried in vacuum for 24 h.

The morphologies of the nanoalloys were analysed using (i) a ZEISS SUPRA 40 field-emission scanning electron microscope and (ii) a JEOL JEM-2100 high resolution transmission electron microscope operated at 200 kV. A drop of the colloidal nanoparticles, pre-sonicated in acetone, was placed on a small Al sheet to prepare the sample for the field-emission scanning electron microscopy (FESEM); the drops were placed on a carbon supported Cu transmission electron microscope grid for the high resolution transmission electron microscopy (HRTEM). To determine the finally syntheiszed composition $x_s$, energy-dispersive X-ray analysis (EDAX) was performed using a JEOL scanning electron microscope. The phases were studied by X-ray diffraction (XRD) using Cu K$_\alpha$ radiation from a Philips X-Pert MRD X-ray diffractometer. X-ray photoelectron spectroscopy (XPS), using a PHI5000 Versaprobe system, was also performed to further verify the stoichiometries of the samples. Micro focussed Al K$\alpha$ (h$\nu$ = 1486.6 eV) X-rays were used for this study, and the binding energy scale was charge referenced to C 1s at 284.5 eV. High-resolution XPS spectra were acquired at 58.7 eV analyser pass energy in steps of 0.25 eV. The study of temperature dependence of resistivities was performed in the range 5 K - 300 K on palletized samples by conventional four-probe method using a homemade resistivity setup with 8 T superconducting magnet from Oxford Instruments Inc., UK at UGC-DAE CSR, Indore. The DC magnetization measurements were performed using the vibrating sample magnetometer of a 14 Tesla Physical property measurement system with 10$^{-5} $ emu sensitivity and 10 mK temperature stability, also at UGC-DAE CSR, Indore.

\section{Results and Discussion}

\subsection{EDAX}

\begin{figure}%
\centering
\includegraphics[width=0.5\textwidth]{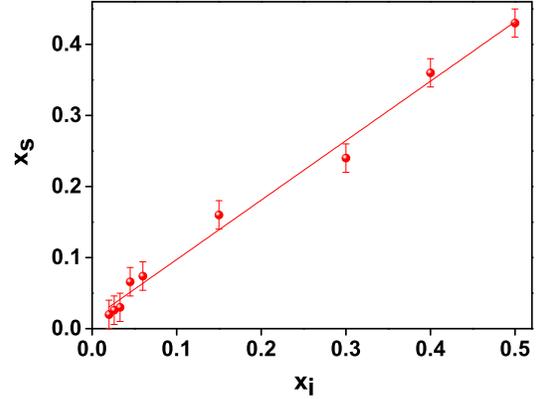}%
\caption{Variation of the composition $x_s$ determined from EDAX and the initial composition $x_i$}
\label{fig:1}%
\end{figure}

Fig. \ref {fig:1} shows the plot between the composition $x_s$ determined from EDAX and the initial composition $x_i$. A linear relation $x_s$ = 0.012 + 0.83 $x_i$ between $x_s$ and $x_i$ is clearly evident from the figure, which provides a scaling between the synthesized and initial Ni concentrations in the samples and confirms that the Ni concentrations taken during the syntheses are almost the same as in the finally synthesized samples. The error bars have been taken to be $\sim$ 2$\%$ of the EDAX determined values as suggested in Ref. \cite{Scott1994}. Here onwards, the value of $x$ in Pd$_{1-x}$Ni$_x$ will be taken as $x_s$.

\subsection{FESEM and HRTEM}

\begin{figure}%
\centering
\subfigure[][]{%
\label{fig:2-a}%
\includegraphics[width=0.22\textwidth]{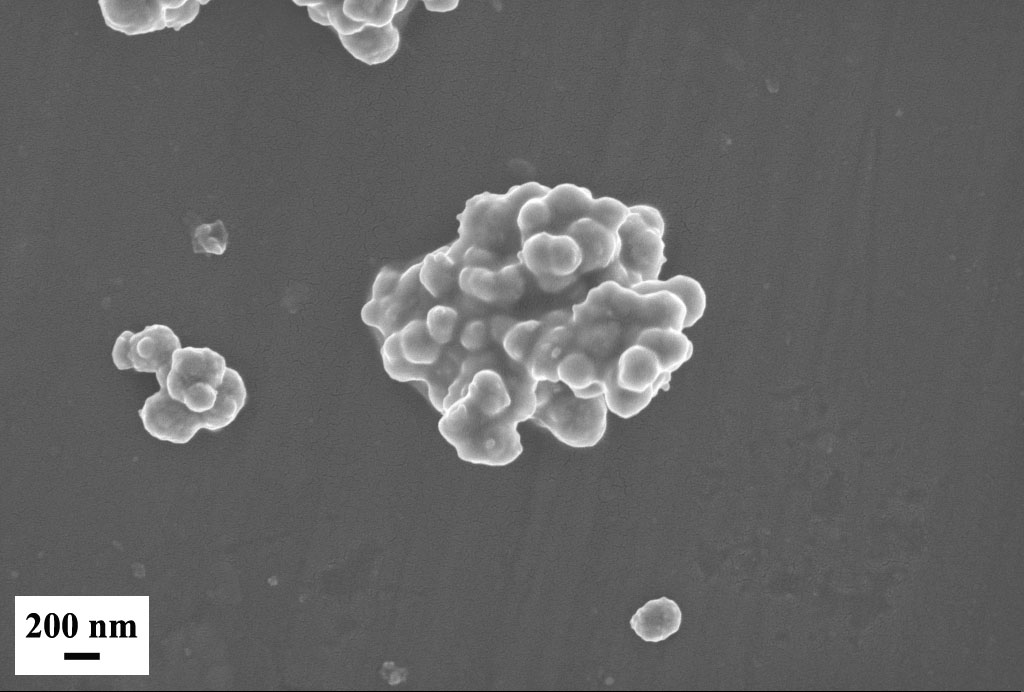}}%
\hspace{8pt}%
\subfigure[][]{%
\label{fig:2-b}%
\includegraphics[width=0.22\textwidth]{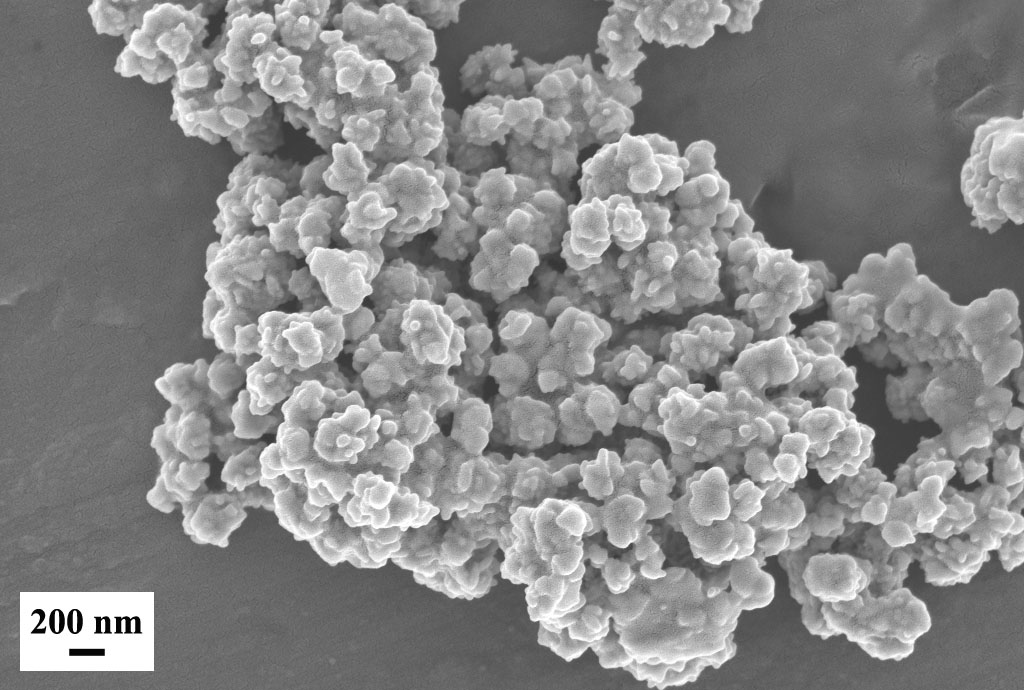}}%
\hspace{8pt}%
\subfigure[][]{%
\label{fig:2-c}%
\includegraphics[width=0.22\textwidth]{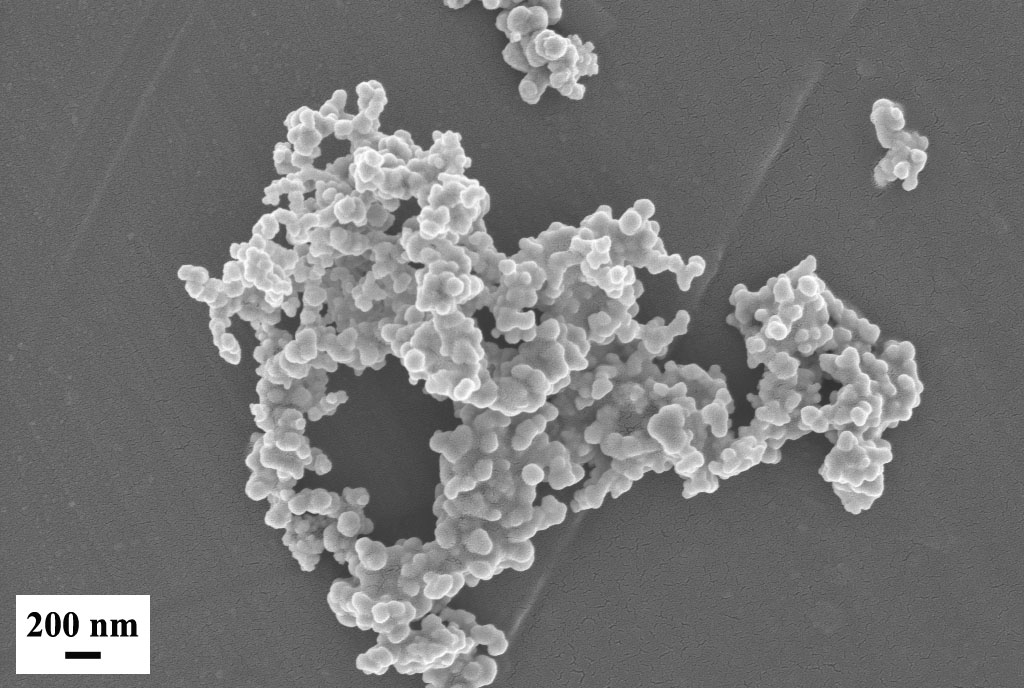}}%
\hspace{8pt}%
\subfigure[][]{%
\label{fig:2-d}%
\includegraphics[width=0.22\textwidth]{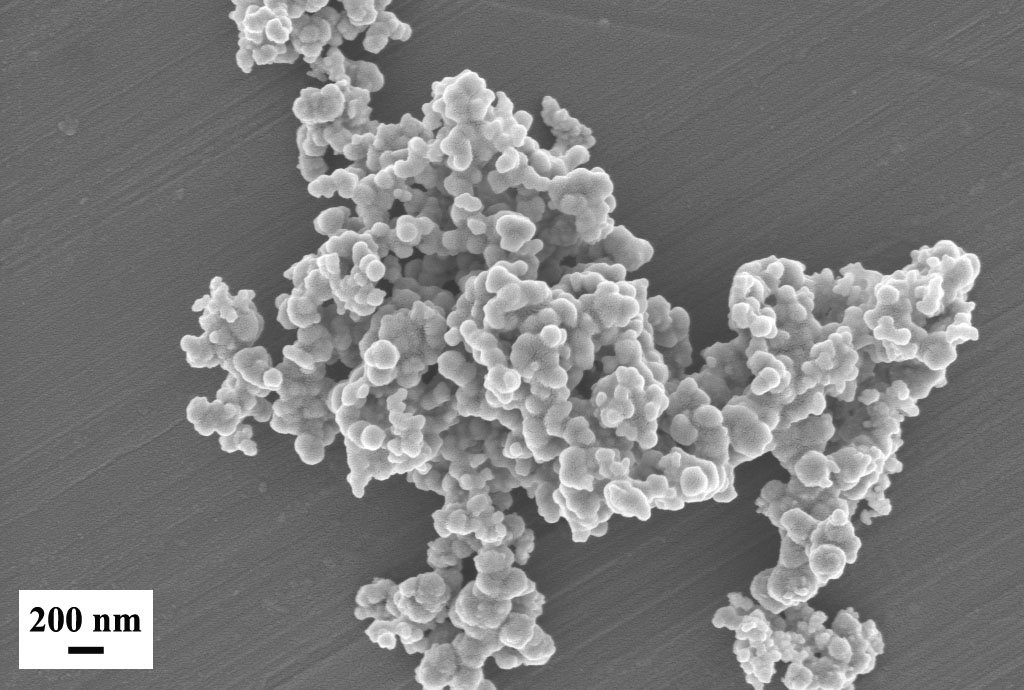}}%
\hspace{8pt}%
\subfigure[][]{%
\label{fig:2-e}%
\includegraphics[width=0.22\textwidth]{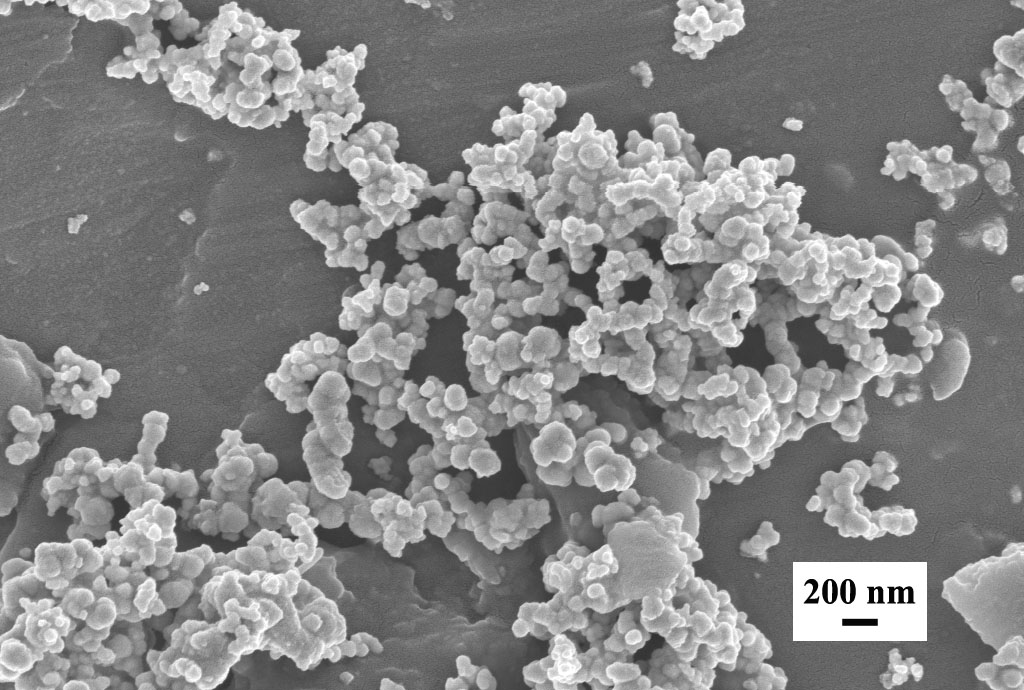}}%
\hspace{8pt}%
\subfigure[][]{%
\label{fig:2-f}%
\includegraphics[width=0.22\textwidth]{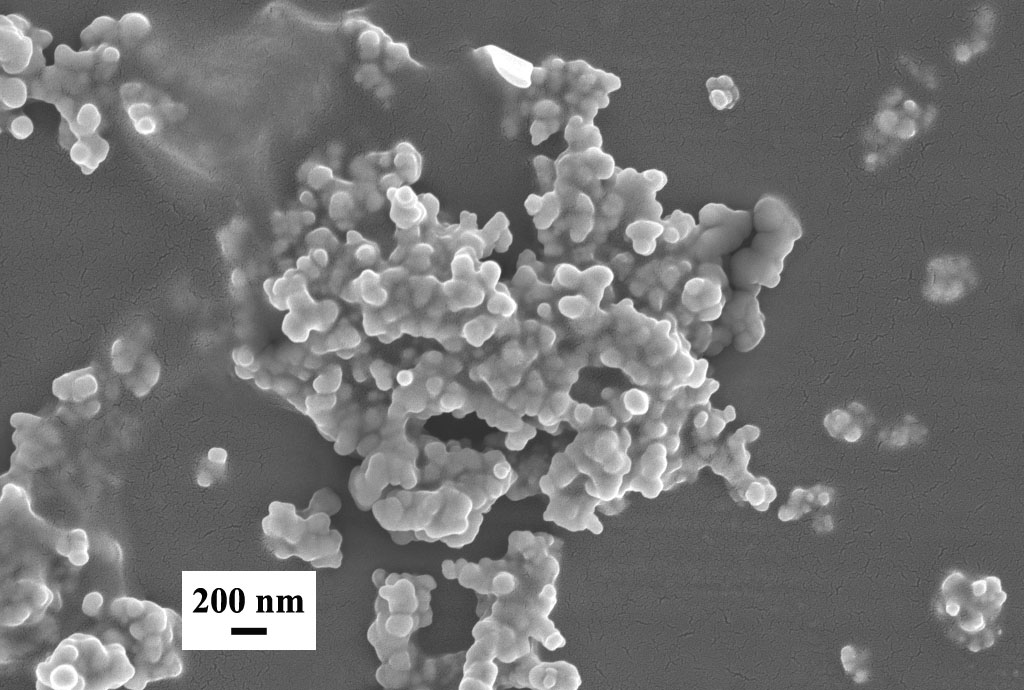}}%
\hspace{8pt}%
\subfigure[][]{%
\label{fig:2-g}%
\includegraphics[width=0.22\textwidth]{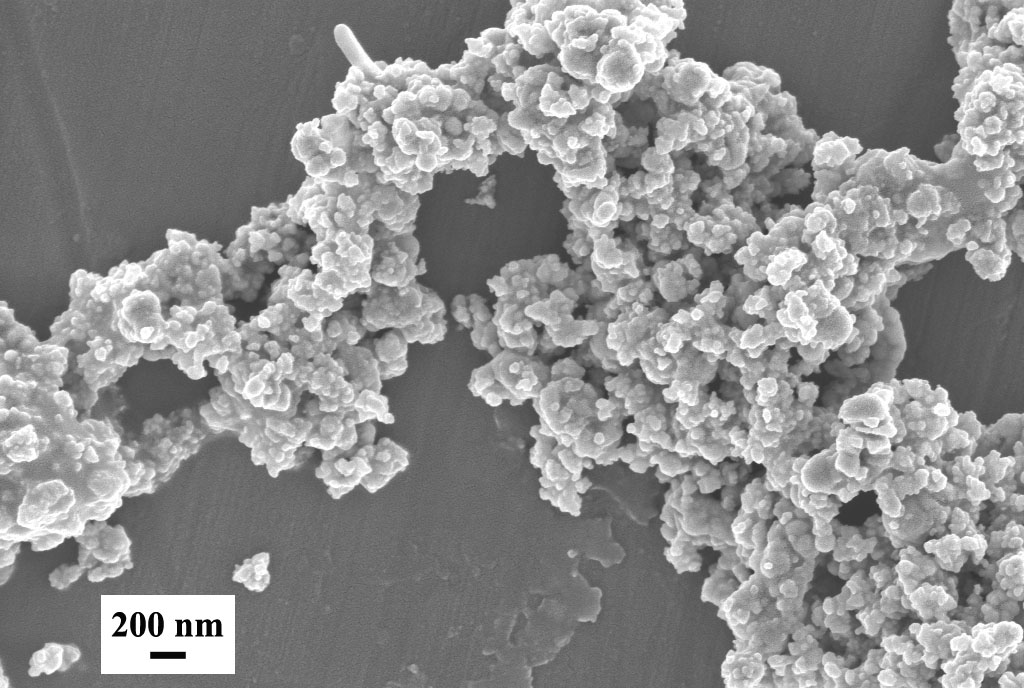}}%
\hspace{8pt}%
\subfigure[][]{%
\label{fig:2-h}%
\includegraphics[width=0.22\textwidth]{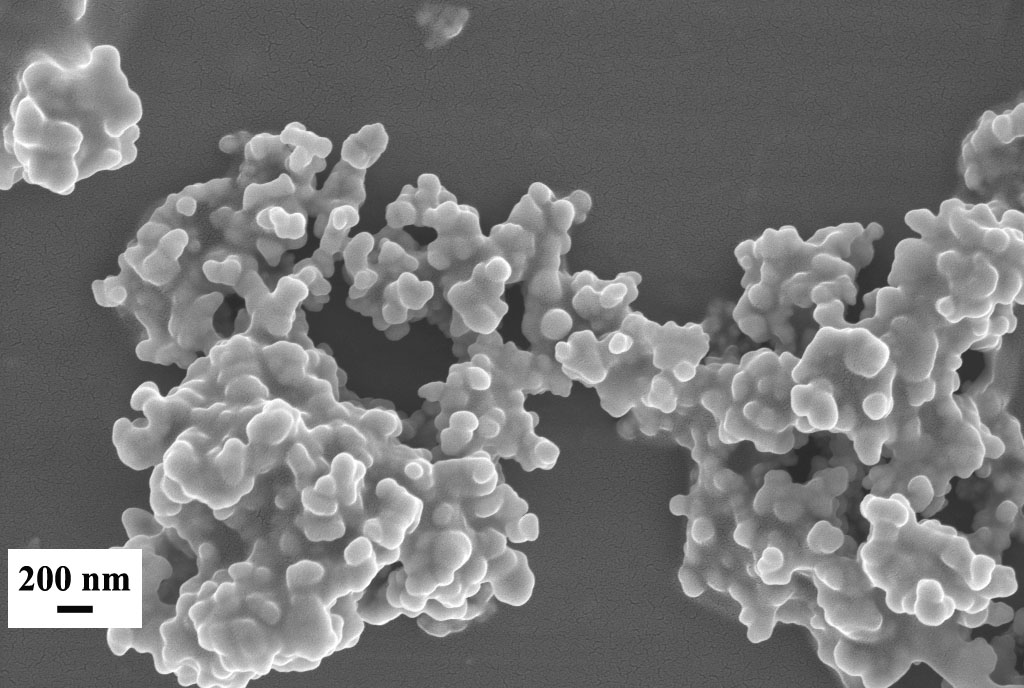}}%

\caption[]{FESEM images of Pd$_{1-x}$Ni$_x$ samples with Ni compositions $x$ =
\subref{fig:2-a} 0.000,
\subref{fig:2-b} 0.020,
\subref{fig:2-c} 0.026,
\subref{fig:2-d} 0.030,
\subref{fig:2-e} 0.074,
\subref{fig:2-f} 0.240,
\subref{fig:2-g} 0.036 and
\subref{fig:2-h} 0.043.}%
\label{fig:2}%
\end{figure}

Fig. \ref {fig:2} shows the FESEM images of the synthesized samples for different Ni concentrations. Ignoring the agglomerations having taken place in the drop before placing it on the Al sheet, formation of 40 - 50 nm diameter spherical nanoparticles can clearly be observed from the figure for all compositions.

\begin{figure}%
\centering
\subfigure[][]{%
\label{fig:3-a}%
\includegraphics[width=0.16\textwidth]{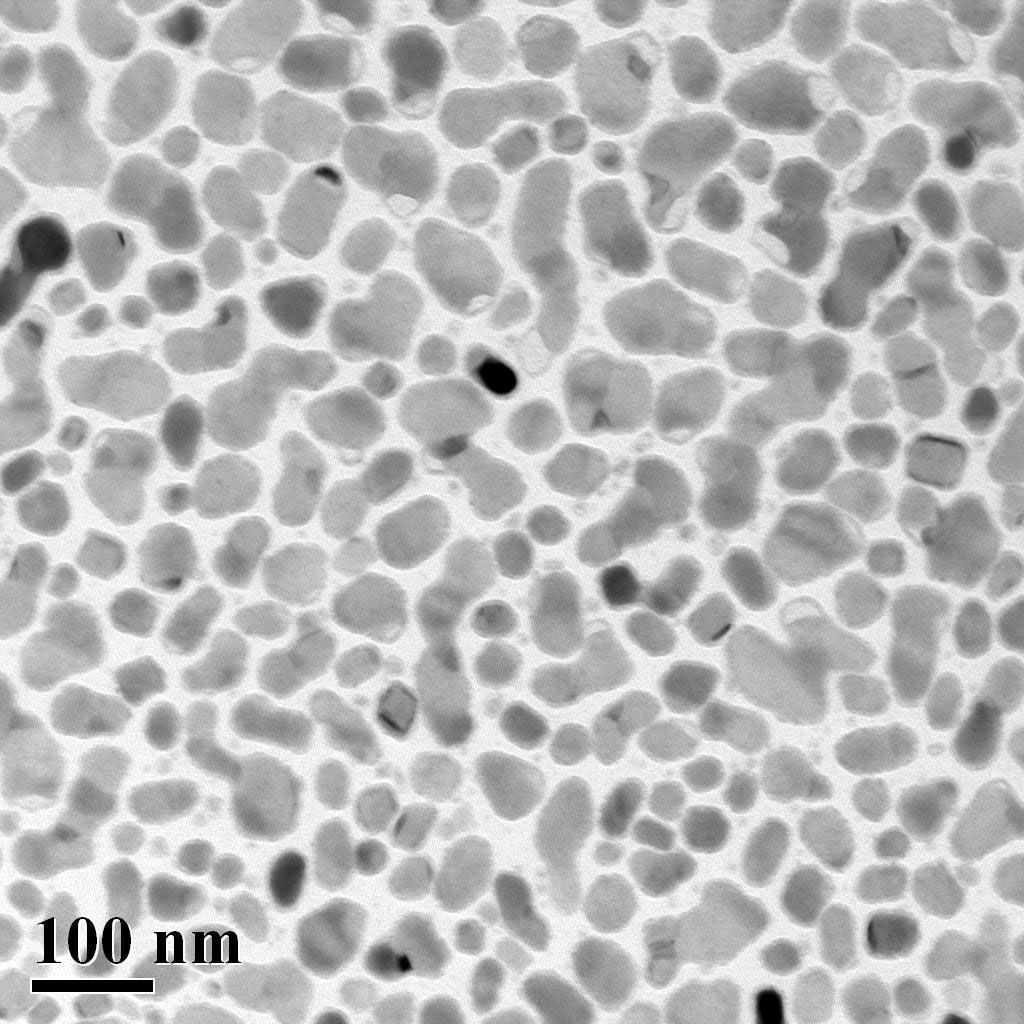}}%
\hspace{8pt}%
\subfigure[][]{%
\label{fig:3-b}%
\includegraphics[width=0.3\textwidth]{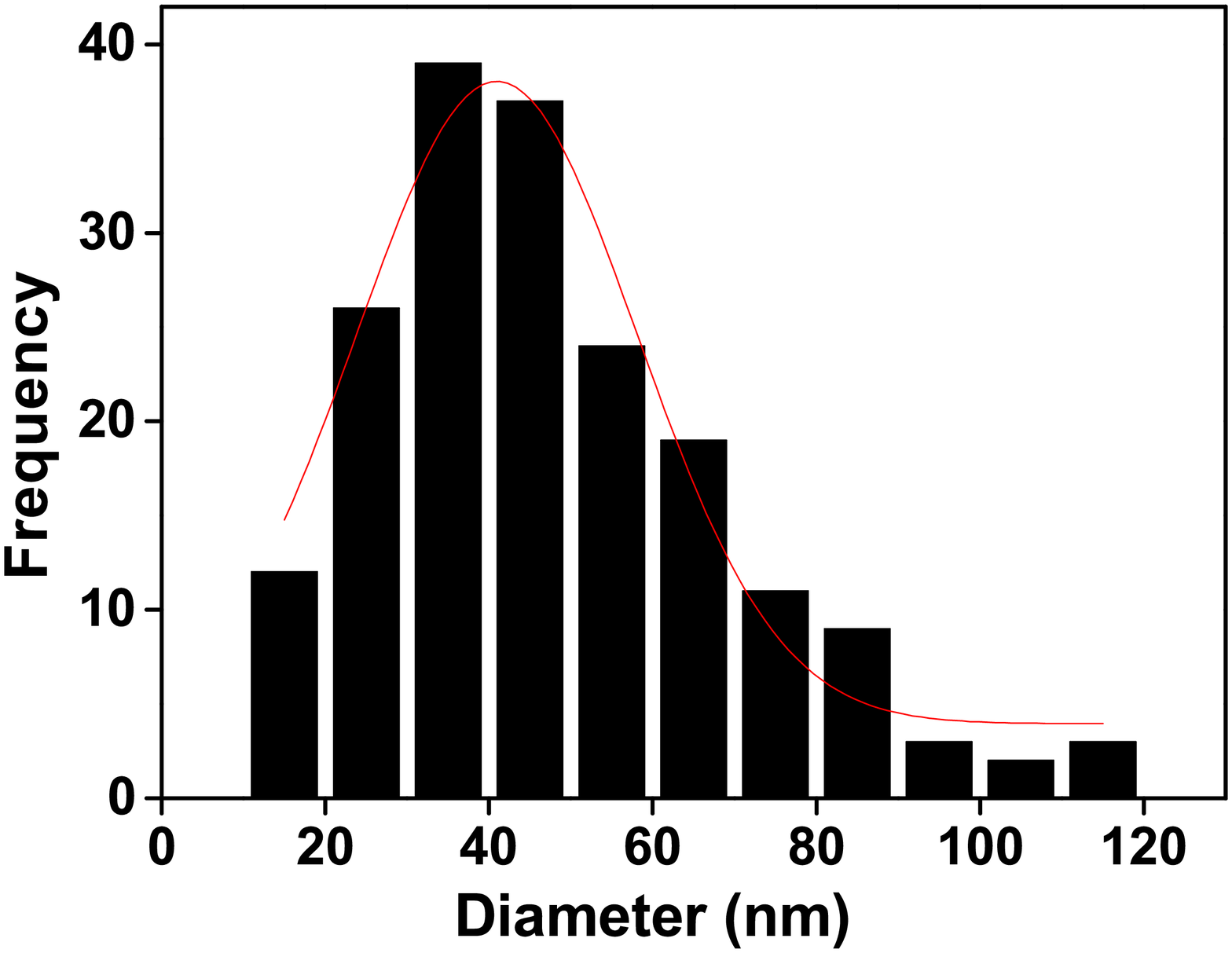}}%
\hspace{8pt}%
\subfigure[][]{%
\label{fig:3-c}%
\includegraphics[width=0.18\textwidth]{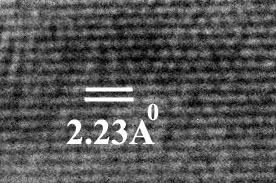}}%
\hspace{8pt}%
\subfigure[][]{%
\label{fig:3-d}%
\includegraphics[width=0.18\textwidth]{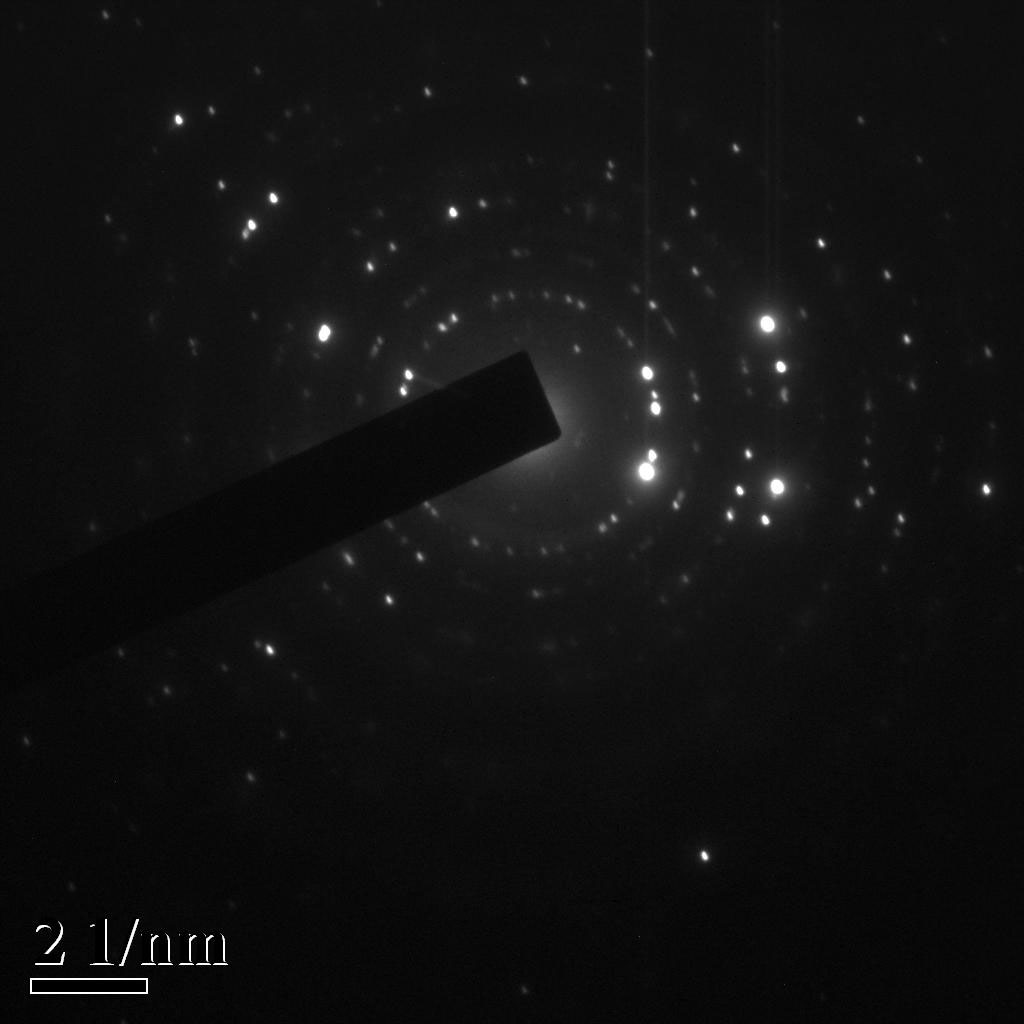}}%
\hspace{8pt}%
\subfigure[][]{%
\label{fig:3-e}%
\includegraphics[width=0.16\textwidth]{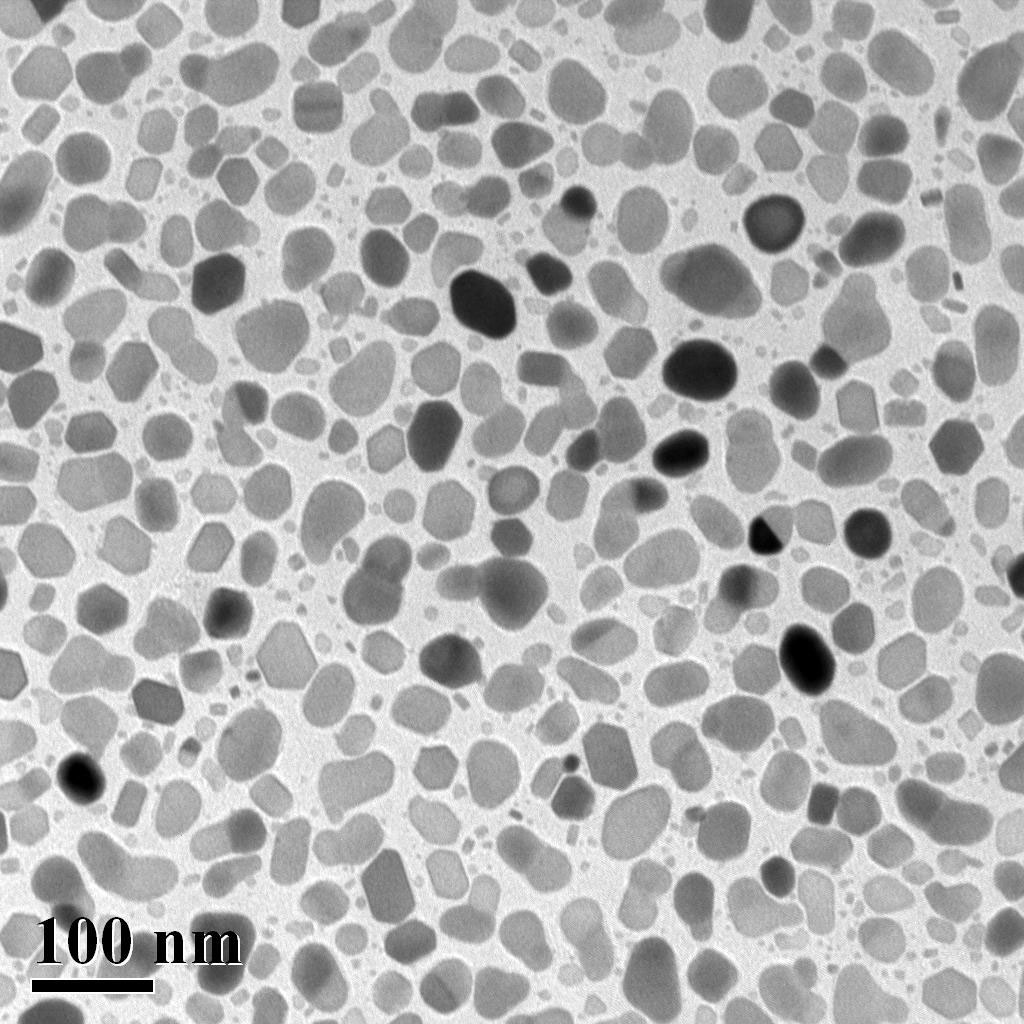}}%
\hspace{8pt}%
\subfigure[][]{%
\label{fig:3-f}%
\includegraphics[width=0.3\textwidth]{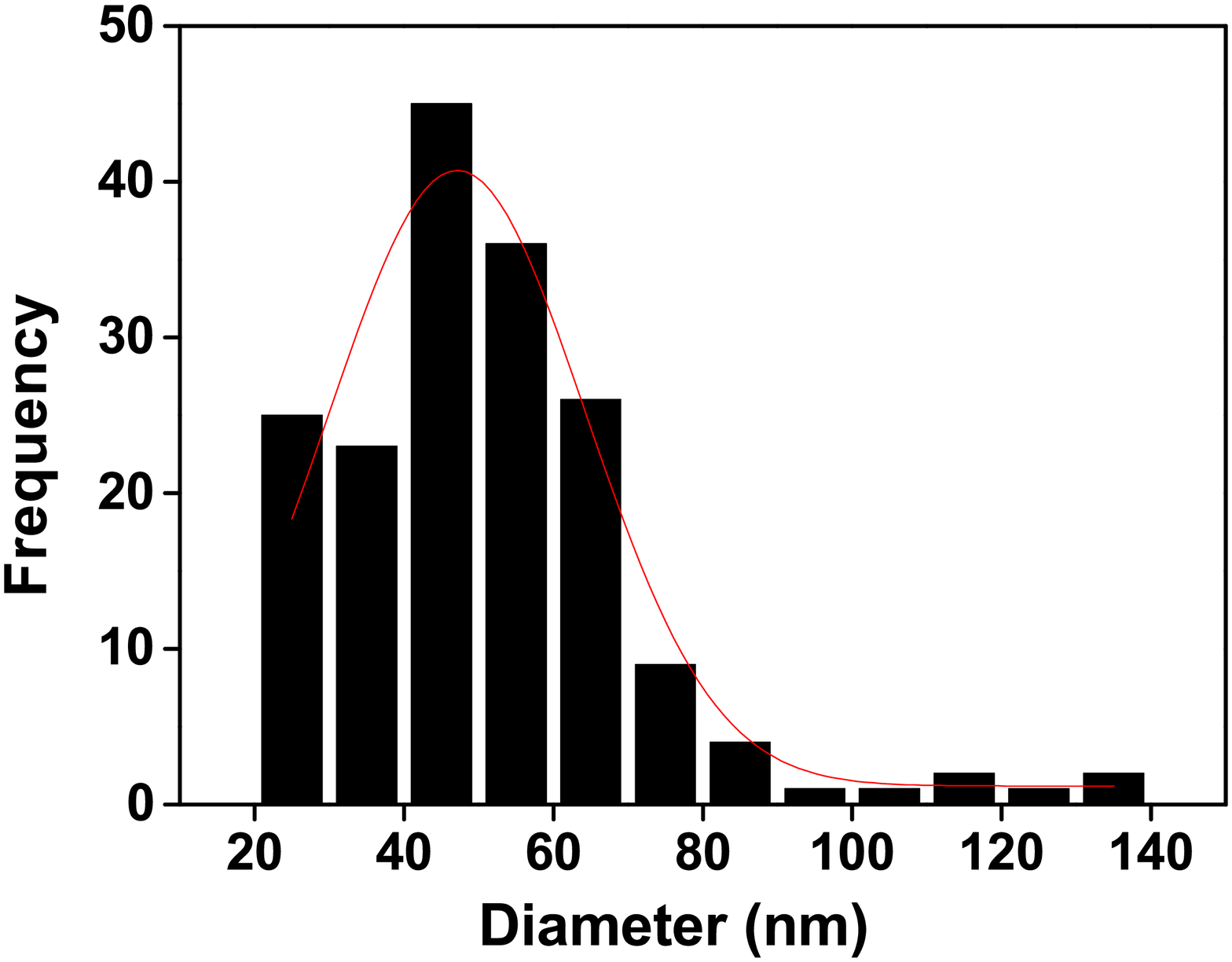}}%
\hspace{8pt}%
\subfigure[][]{%
\label{fig:3-g}%
\includegraphics[width=0.2\textwidth]{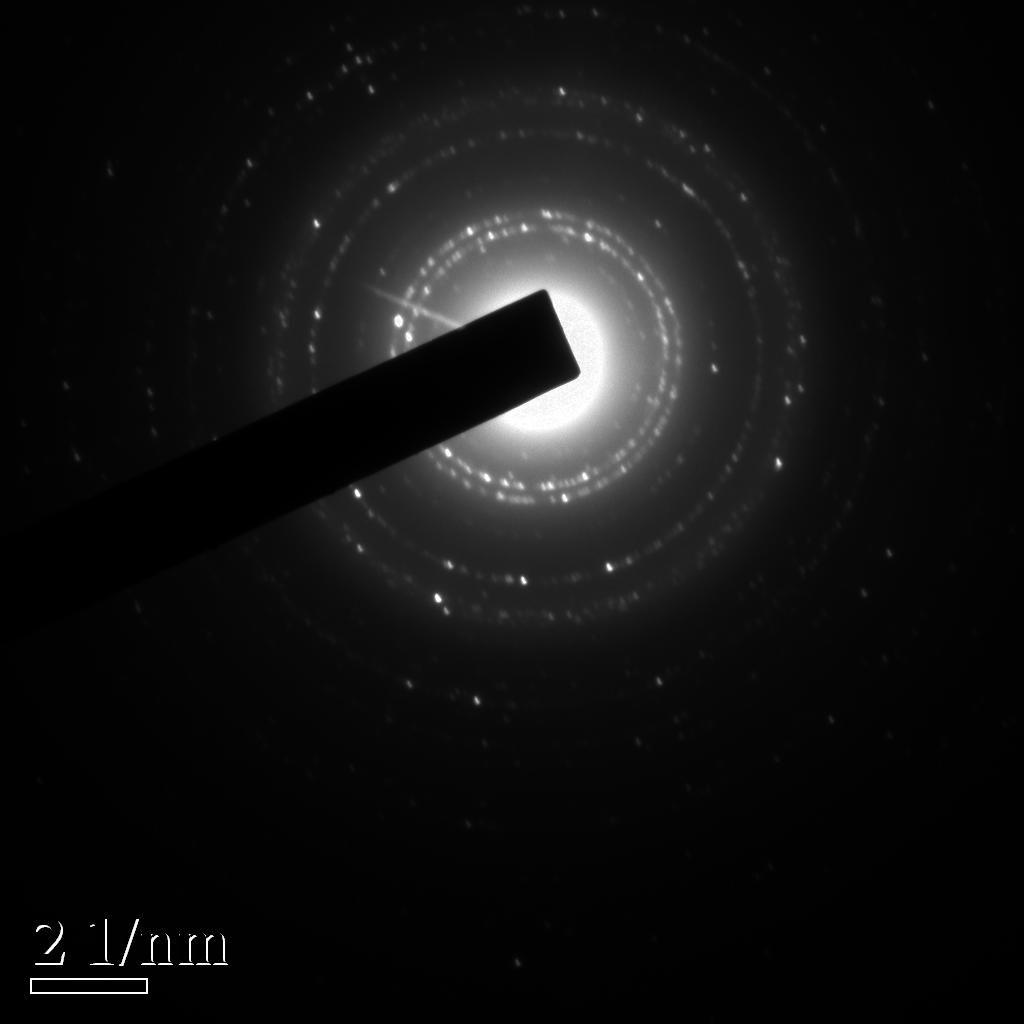}}%

\caption[]{HRTEM micrographs of Pd$_{1-x}$Ni$_x$ samples:
\subref{fig:3-a} the HRTEM image for $x$ = 0.036,
\subref{fig:3-b} the particle size distribution for $x$ = 0.036,
\subref{fig:3-c} the selected area diffraction pattern of $x$ = 0.036,
\subref{fig:3-d} a higher resolution image of $x$ = 0.036 showing the lattice planes and crystallinity,
\subref{fig:3-e} the HRTEM image for $x$ = 0.043,
\subref{fig:3-f} the particle size distribution for $x$ = 0.036, and
\subref{fig:3-g} the selected area diffraction pattern of $x$ = 0.036.}%
\label{fig:3}%
\end{figure}

For a better visualization of the nanoparticles, HRTEM images were taken for two representative samples with $x$ = 0.36 and 0.43. The image for $x$ = 0.36 is shown in Fig. \ref{fig:3-a}. From the image, formation of faceted nanoparticles of $\sim$ 40-50 nm sized particles or their elongations due to coalescence of such particles is clearly observed. Faceting of fcc metal nanoparticles, like the one observed here, is in agreement with a report by Karkina {\it et al.} \cite{Karkina2010} via a molecular dynamics simulation. The  particle size distribution for this sample, as shown in Fig. \ref{fig:3-b}, confirms that the mean size of the nanoparticles is $\sim$ 40-50 nm; the larger sizes, as discussed already, are due to coalescence induced elongations of the particles. The bright spots along with the concentric rings as shown in the selected area diffraction (SAED) pattern, Fig. \ref{fig:3-c}, for this sample is indicative of formation of crystalline nanoparticles with random orientations of smaller domains inside. An even higher resolution image of a particle in this sample, as shown in Fig. \ref{fig:3-d} confirms the crystallinity and measures the interplanar spacing to be 2.23 {\AA}, which is between the (111) interplanar spacings of Pd (2.25 {\AA}) and Ni (2.03 {\AA}) and indicates the alloy formation between Pd and Ni. The figure also suggests the formation of single phase structure of the nanoparticles without any core-shell feature. The same conclusions, except for the lattice spacings, can be drawn for the sample with $x$ = 0.43 from its HRTEM image, Fig. \ref{fig:3-e}, the particle size distribution, Fig. \ref{fig:3-f}, and the SAED pattern, Fig. \ref{fig:3-g}. With the high-resolution imaging of these two representative samples, and with the visual uniformity of the FESEM images of all the samples, we can conclude that all the synthesized samples are of $\sim$ 40-50 nm size along with a few elongated particles and are crystalline nanoalloys.

\subsection{XRD}

\begin{figure}%
\centering
\subfigure[][]{%
\label{fig:4-a}%
\includegraphics[width=0.2\textwidth]{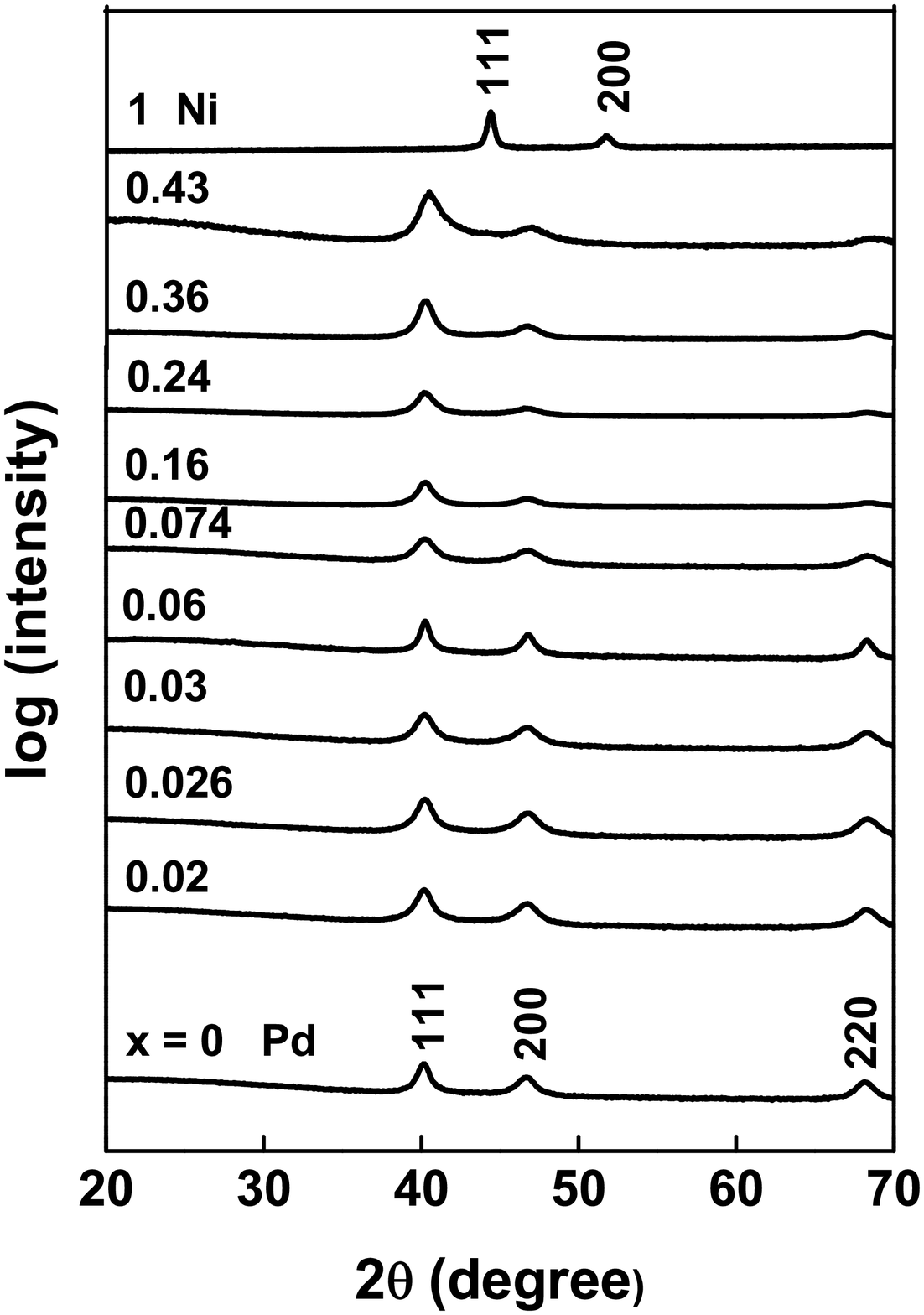}}%
\hspace{8pt}%
\subfigure[][]{%
\label{fig:4-b}%
\includegraphics[width=0.22\textwidth]{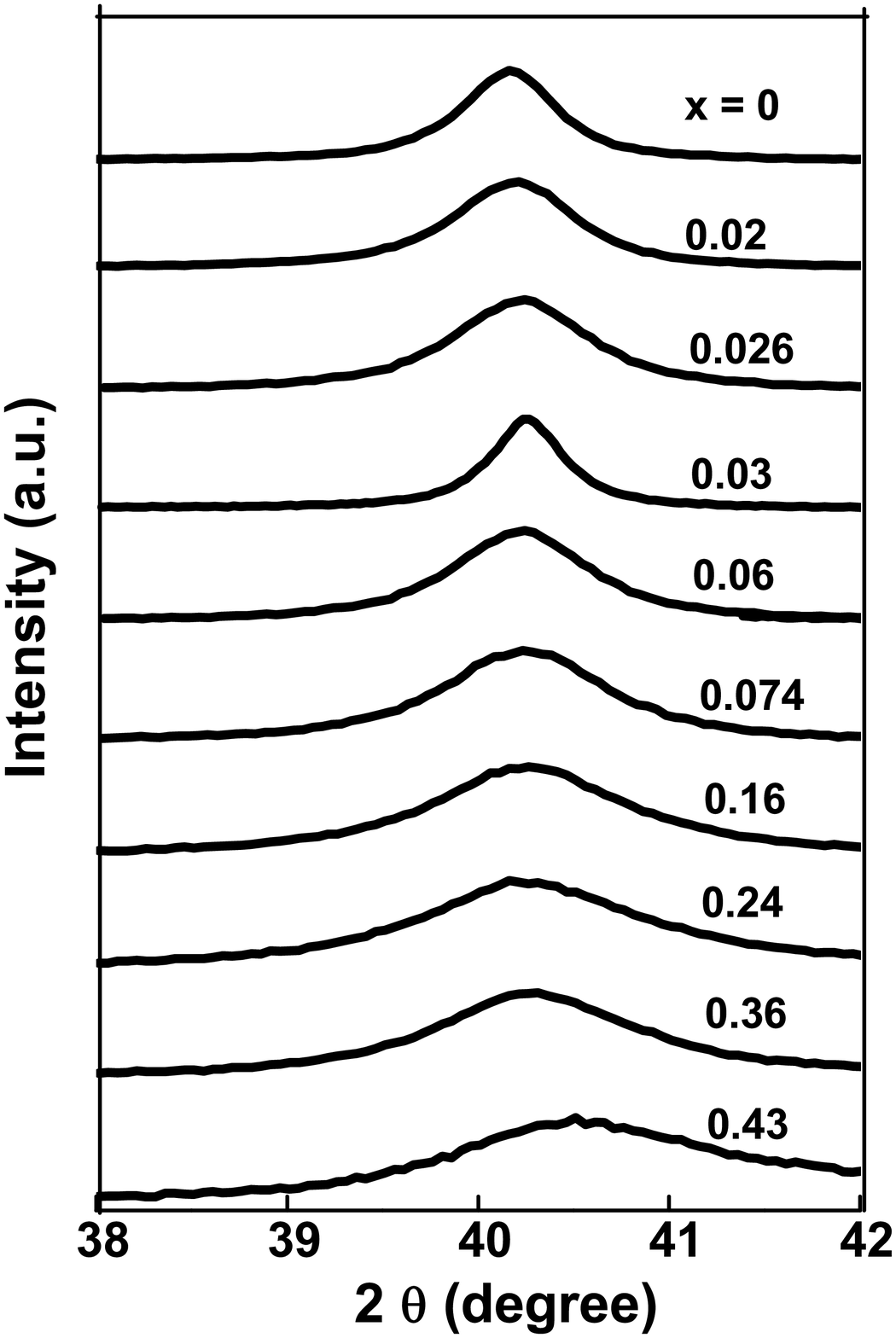}}%
\hspace{8pt}%
\subfigure[][]{%
\label{fig:4-c}%
\includegraphics[width=0.35\textwidth]{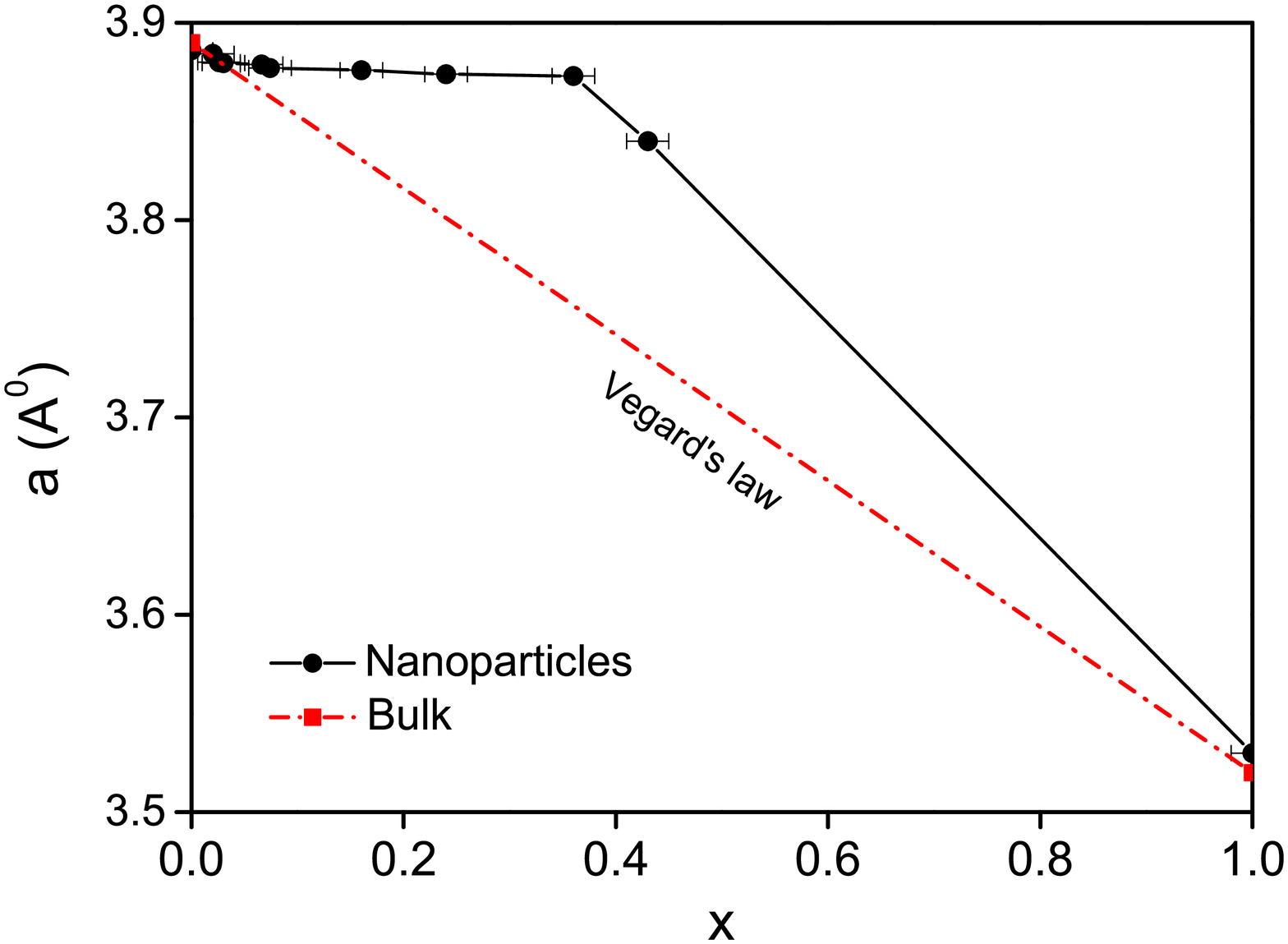}}%

\caption[]{
\subref{fig:4-a} XRD patterns of Pd$_{1-x}$Ni$_x$ samples.
\subref{fig:4-b} Magnified view of (111) peaks.
\subref{fig:4-c} Variation of lattice constant $a$ of the synthesized alloy with $x$.}
\label{fig:4}%
\end{figure}

The XRD patterns for all the spectra, including the XRD spectrum of pure Ni ($x$ = 1) nanoparticles prepared in the same manner as others, are shown in Fig. \ref{fig:4-a}. Sharp and strong reflection peaks at 2$\theta$ values of $40.10^o$, $46.64^0$, and $68.10^o$, as verified from the JCPDS data, correspond to the (111), (200), and (220) planes of the fcc crystallographic structure of Pd. The sharpness of the peaks corroborates the inferences from the HRTEM images about the good crystallinity of the samples. In addition, the absence of all Ni peaks (observed for the pure Ni nanoparticles) in the XRD patterns of all the alloy nanoparticles suggests a complete alloying of Pd and Ni for all $x$ values under study. Fig. \ref{fig:4-b} shows a magnified view of the XRD patterns in the vicinity of the (111) peaks. The observed composition dependent systematic shift in the peak position confirms the alloying of Pd and Ni with varying Ni concentrations. Further, the lattice constants have been determined from all XRD patterns and are plotted as a function of $x$ in Fig. \ref{fig:4-c}. Although a marked deviation of the lattice constants from the Vegard's law is visible from the figure, such deviations have already been reported for PdNi nanoparticles \cite{Bagaria}. Furthermore, the lattice constant for $x$ = 0.036 comes out to be 3.87 {\AA} as deduced from the corresponding XRD pattern. This leads to (111) lattice spacing of 2.24 {\AA} in Pd$_{0.74}$Ni$_{0.36}$, in excellent agreement with the HRTEM results.

\subsection{XPS}

\begin{figure}%
\centering
\subfigure[][]{%
\label{fig:5-a}%
\includegraphics[width=0.35\textwidth]{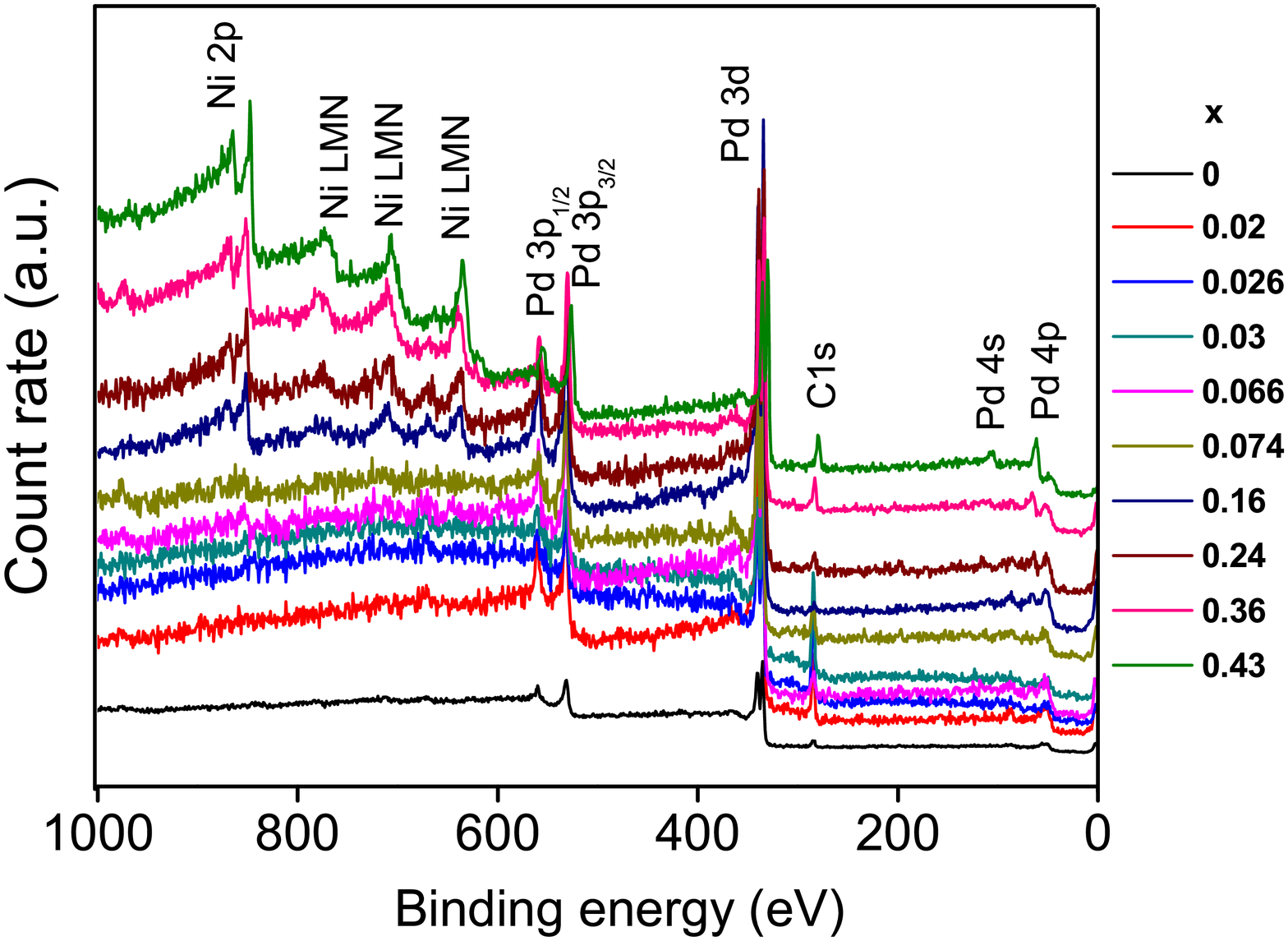}}%
\hspace{8pt}%
\subfigure[][]{%
\label{fig:5-b}%
\includegraphics[width=0.22\textwidth]{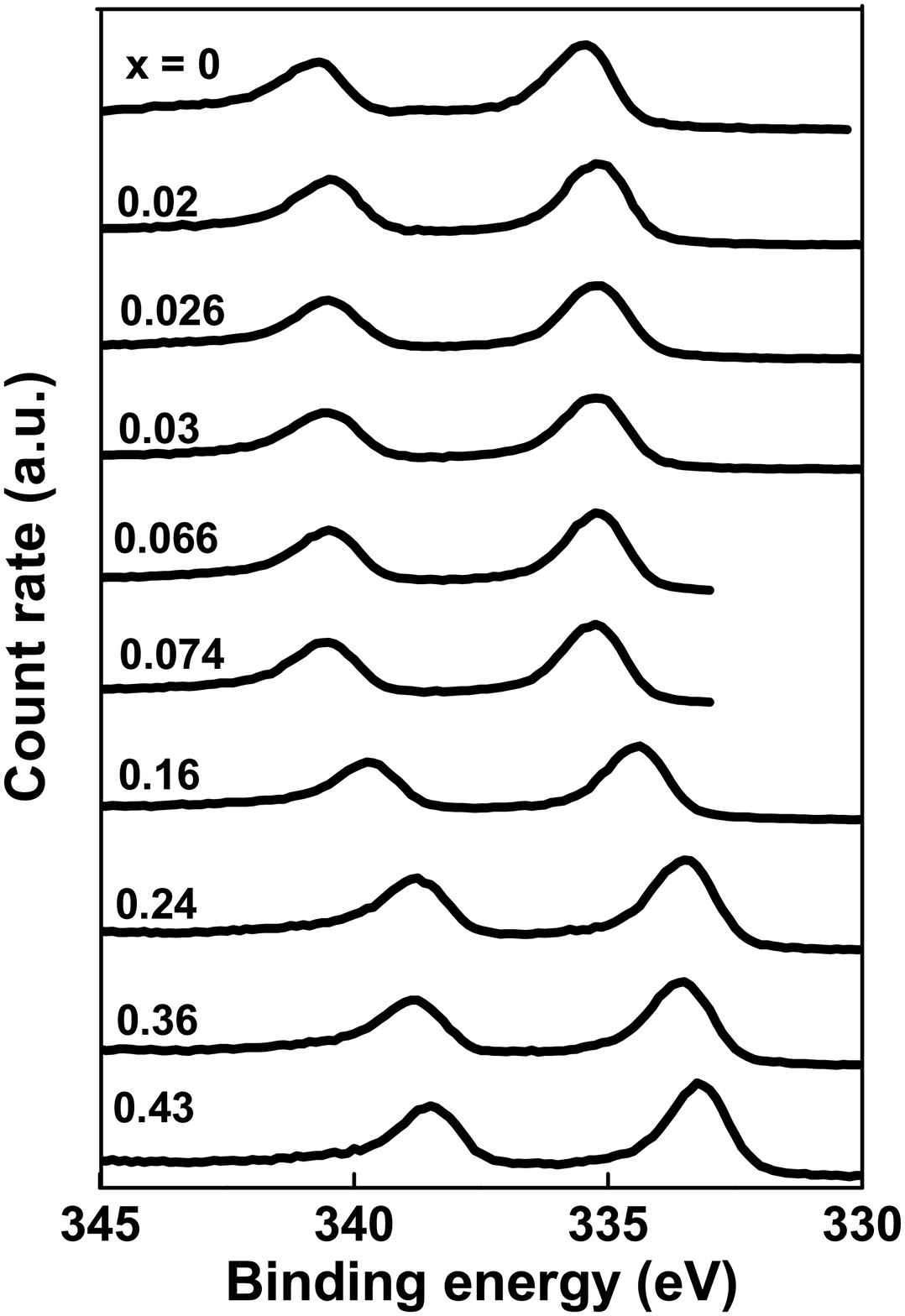}}%
\hspace{8pt}%
\subfigure[][]{%
\label{fig:5-c}%
\includegraphics[width=0.22\textwidth]{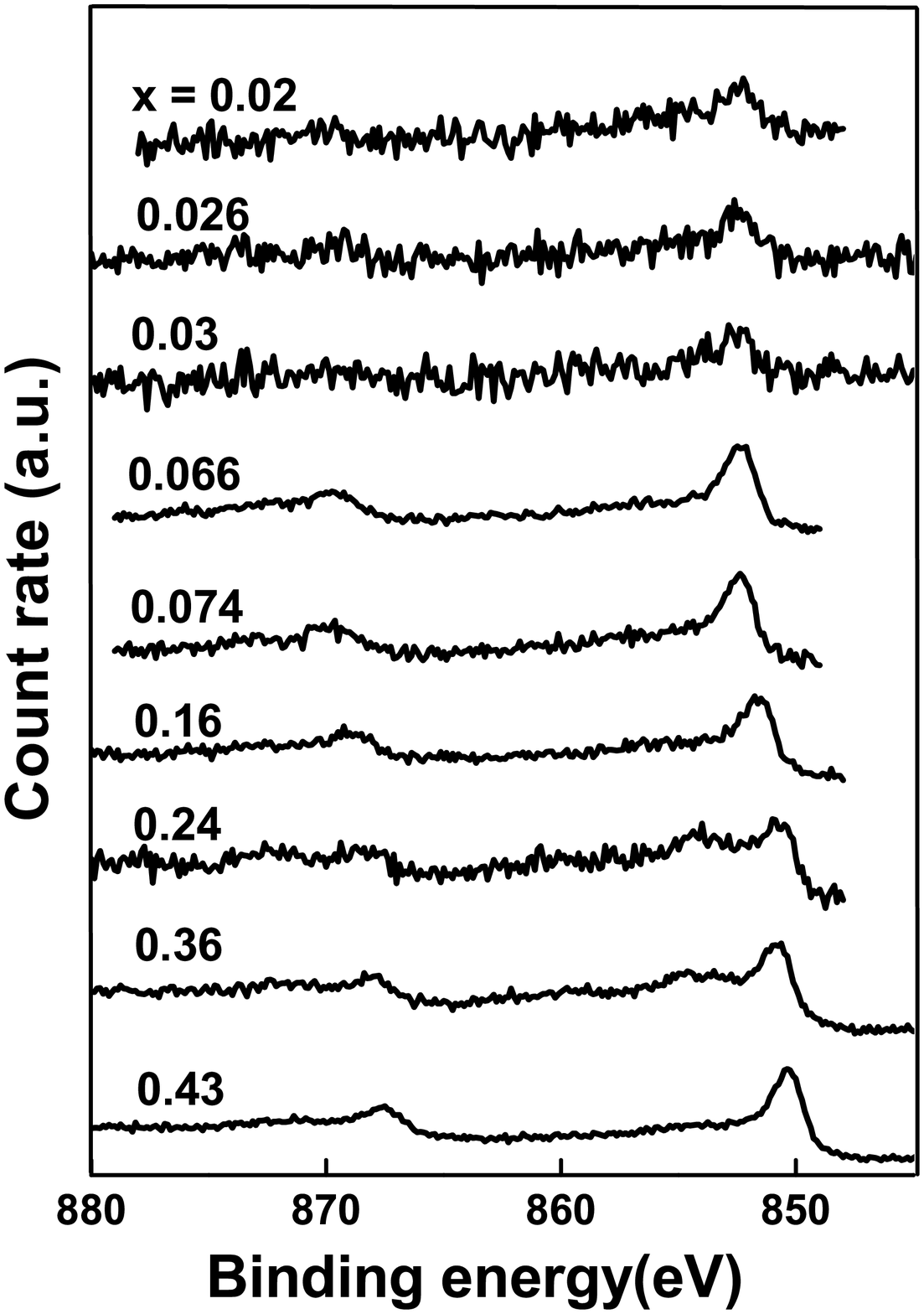}}%
\hspace{8pt}%
\subfigure[][]{%
\label{fig:5-d}%
\includegraphics[width=0.35\textwidth]{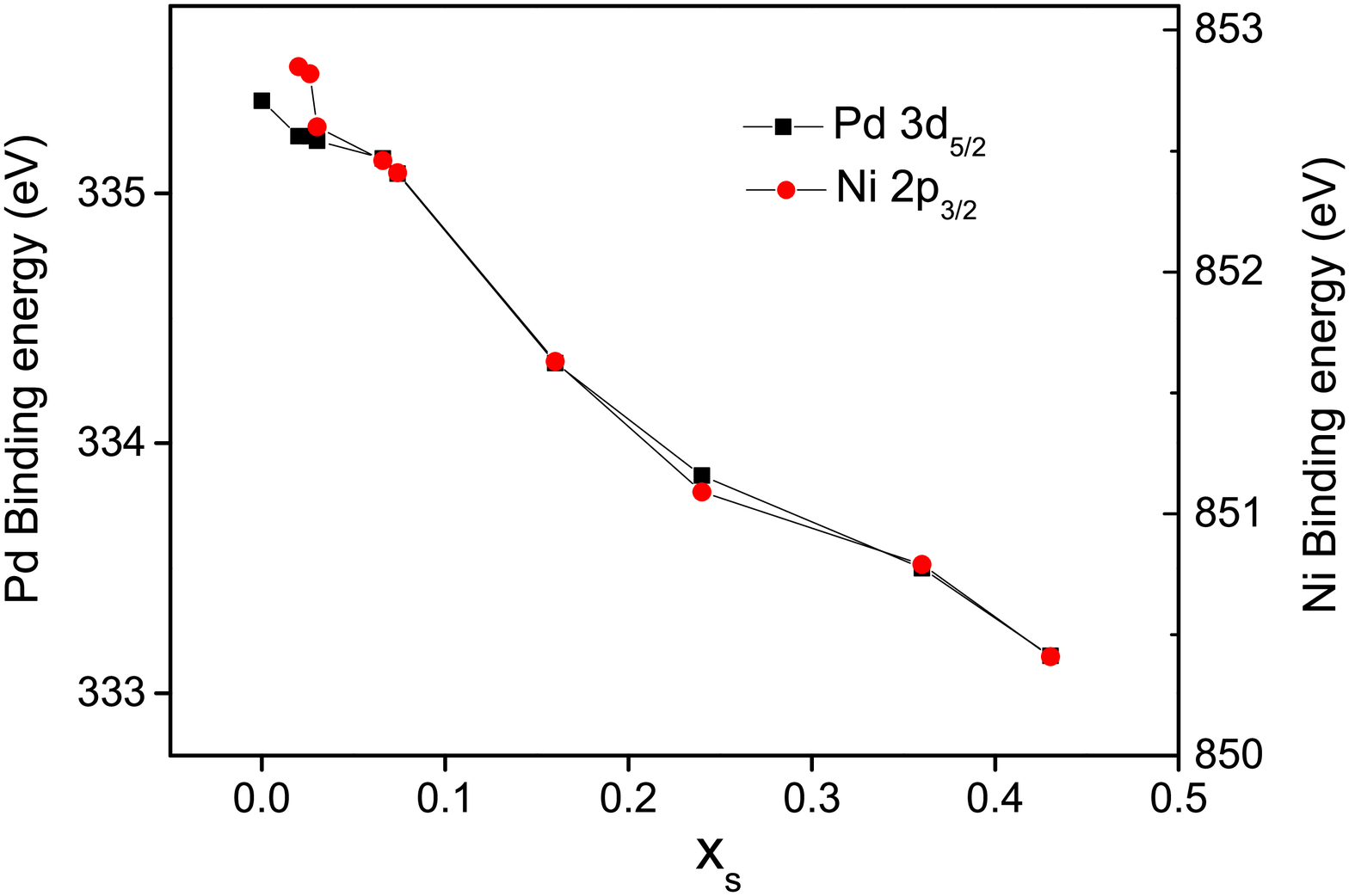}}%

\caption[]{
\subref{fig:5-a} Survey XPS spectra of Pd$_{1-x}$Ni$_x$ samples.
\subref{fig:5-b} High-resolution XPS spectra in the Pd 3d region.
\subref{fig:5-c} High-resolution XPS spectra in the Ni 2p region.
\subref{fig:5-d} Variation of Pd and Ni peak position with $x$.}
\label{fig:5}%
\end{figure}

For an additional verification of the alloy and phase formation, the samples were analyzed by XPS. The survey spectra of all the samples, as displayed in Fig. \ref{fig:5-a}, reveal the presence of both Pd and Ni in all the samples. The peak at 284.5 eV is the C 1s peak coming due to an unavoidable presence of hydrocarbons on the sample surface; all the XPS spectra, as mentioned earlier, have been charge referenced to this peak. There could be a contribution to the pure Pd 3p$_{3/2}$ peak occurring at around 532.4 eV \cite{Militello1994_1} from O 1s occurring at 530.5 eV due to PdO \cite{Militello1994_2}. So, we need to analyze a region of the XPS spectra containing, e.g., Pd 3d peaks in more detail to look for any other peak occurring due to PdO in that region. Fig. \ref{fig:5-b} displays high-resolution XPS spectra for all the samples in the Pd 3d region. PdO, if present, would have manifested itself as an O 1s peak at 336.8 eV \cite{Brun1999}. Its absence in all the spectra reveals that there is no palladium oxide present in any sample. To further explore the possibility of the presence of any nickel oxide, high-resolution spectra in the Ni 2p region were also recorded and are presented in Fig. \ref{fig:5-c}. One would expect a NiO or Ni(OH)$_2$ peak between 853.7 and 855.6 eV. This region of the spectra does not seem to have any observable peak structure, ruling out the presence of even oxides of nickel. The samples are, thus, essentially oxide-free and are suitable for further analysis. It is to be noted that the Pd 3d$_{5/2}$ peak occurring at 335.4 eV for the synthesized pure Pd ($x$ = 0) nanoparticles is shifted by +0.37 eV with respect to bulk Pd value. Such a shift, however, is expected for nanoparticles \cite{Shukla1999} and hence corroborates the FESEM and HRTEM images.

Further, composition dependent shifts of Pd 3d$_{5/2}$ (335.4 eV - 333.2 eV) and Ni 2p$_{3/2}$ peaks (853.6 eV - 850.4) are also observable from Figs. \ref{fig:5-b} and \ref{fig:5-c}, respectively. Plots of the Pd 3d$_{5/2}$ and Ni 2p$_{3/2}$ peak positions with the composition $x$ are shown in Fig. \ref{fig:5-d}. If we ignore the low-$x$ points corresponding to Ni 2p$_{3/2}$ peaks because of their determination due to low counts at these concentrations, both the peaks shift to lower binding energies in a somewhat linear fashion due to alloying \cite{Steiner1981}. This confirms that the synthesized nanoparticles are of alloys of the EDAX determined compositions and justifies the deviation of the lattice constants from the Vegard's law.

\subsection{Resistivity}

\begin{figure}%
\centering
\subfigure[][]{%
\label{fig:6-a}%
\includegraphics[width=0.35\textwidth]{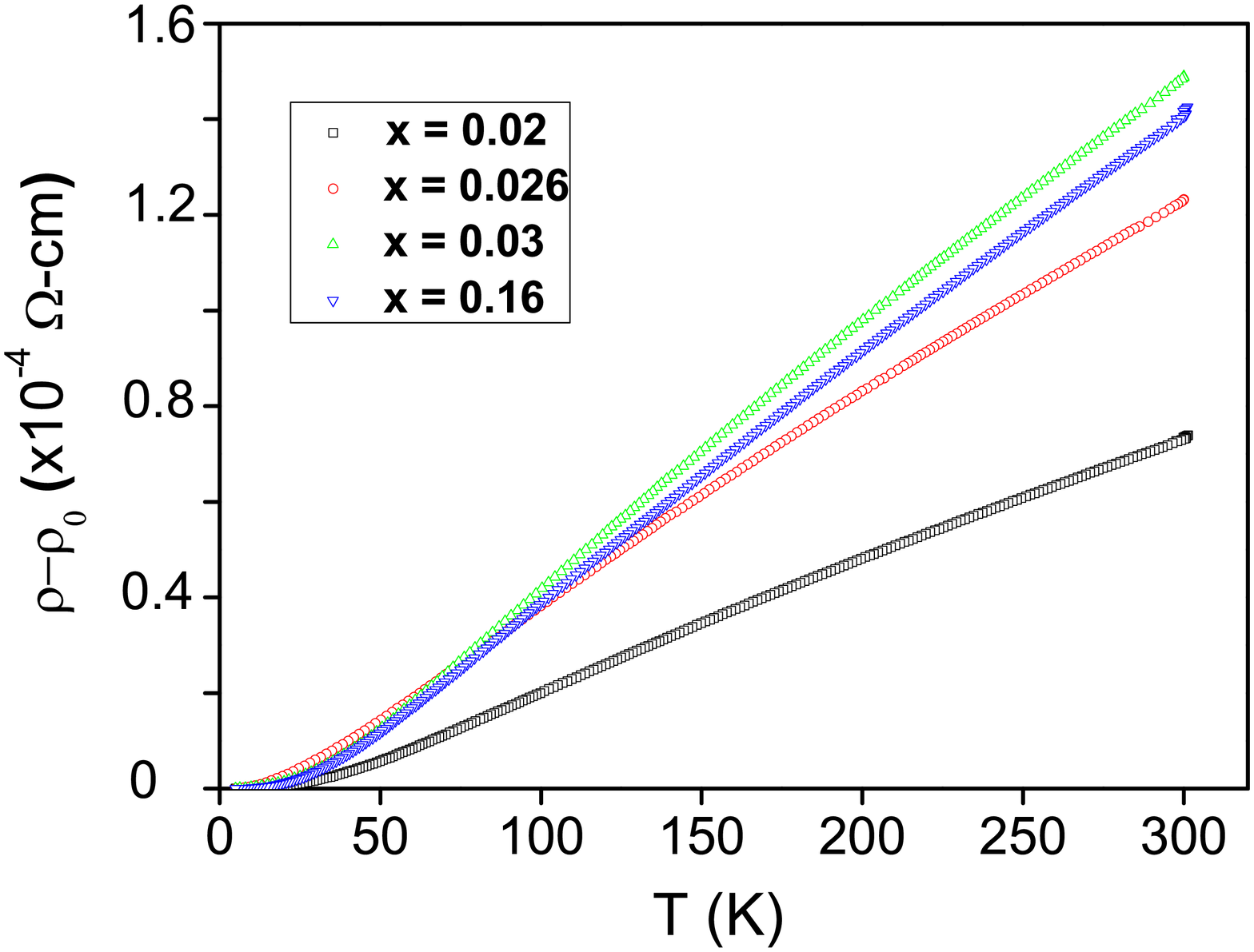}}%
\hspace{8pt}%
\subfigure[][]{%
\label{fig:6-b}%
\includegraphics[width=0.35\textwidth]{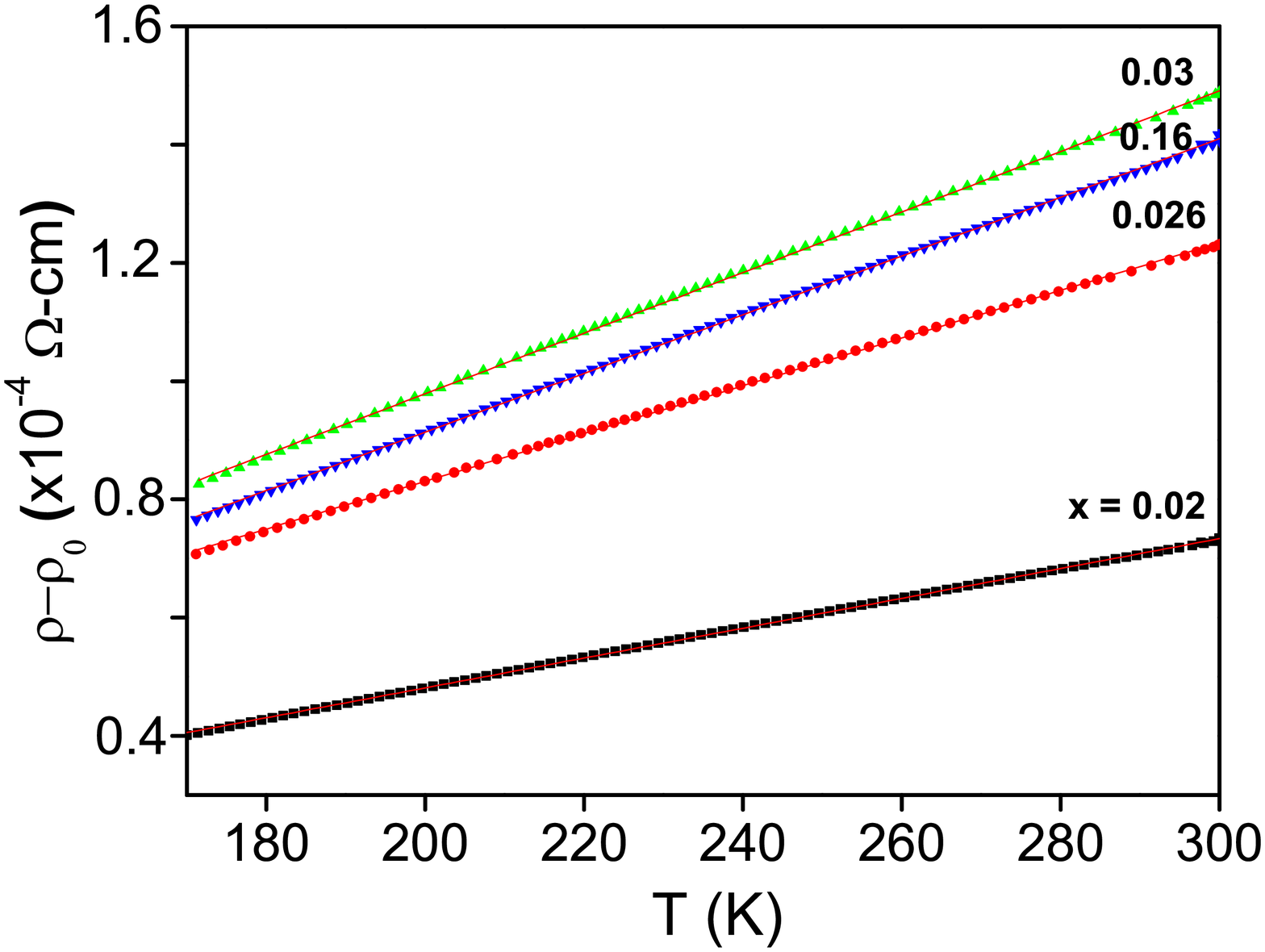}}%
\hspace{8pt}%
\subfigure[][]{%
\label{fig:6-c}%
\includegraphics[width=0.35\textwidth]{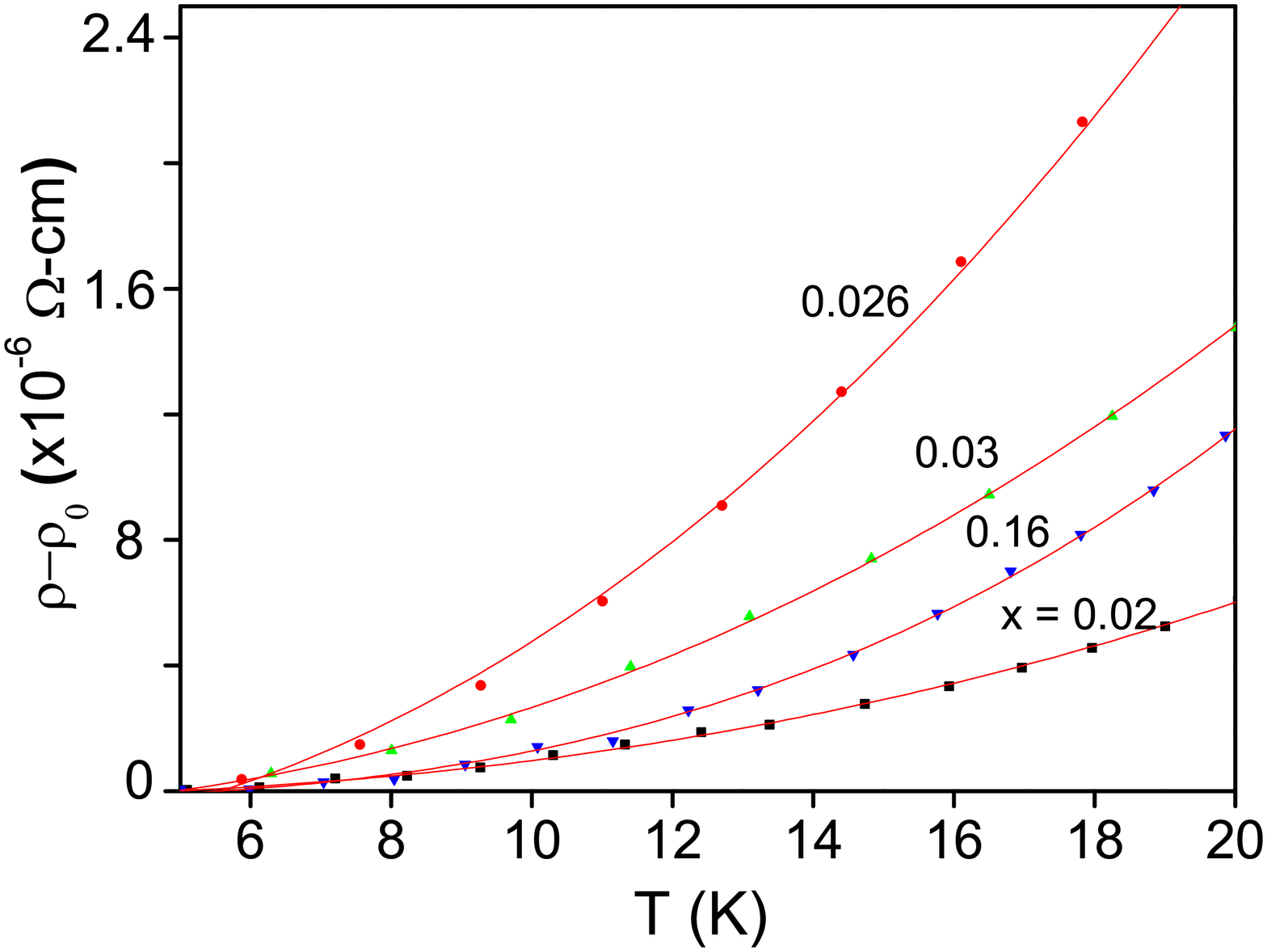}}%

\caption[]{Temperature dependence of resistivity
\subref{fig:6-a} Full temperature range.
\subref{fig:6-b} High-T range with linear fits.
\subref{fig:6-c} Low-T (2-20 K) range with $AT^n$ fits.}
\label{fig:6}%
\end{figure}

Measurements of the temperature $T$ dependence of resistivity $\rho$ were performed on four representative nanoalloys with $x$ = 0.02, 0.026, and 0.03 on the lower Ni concentration side, and $x$ = 0.16 on the higher Ni concentration side. The residual resistivity $\rho_0$ has been subtracted from the measured $\rho$ in order to plot the data. Fig. \ref{fig:6-a} shows the plots of $\rho - \rho_0$ versus $T$ for these samples in the temperature interval 5 K - 300 K. The metallic nature of the nanoalloys is clearly confirmed by the monotonic increase of $\rho - \rho_0$ with $T$ for all the samples. The high-T (170 K - 300 K) resistvity (Fig. \ref{fig:6-b}) shows a linear variation with T, arising due to the dominant electron-phonon interactions in this temperature range \cite{White59}. Fig. \ref{fig:6-c} shows the low temperature (5 K - 20 K) part of the resistvities for the four samples and their power law $\rho(T)-\rho_0 = AT^n$ fits, where $A$ and $n$ are generalized Fermi liquid (FL) coefficient and generalized temperature exponent, respectively, as defined by Nicklas $\it et ~al.$ \cite{Nicklas99}. The values of $A$ and $n$ are shown in Tab. \ref{tab1}. The decrease in the exponent from 2.9$\pm$0.04 at $x$ = 0.16 to 2.1$\pm$0.06 for $x$ = 0.026 and a minor upturn to the value 2.2$\pm$0.04 at $x$ = 0.02 are in agreement with the report by Nicklas $\it et ~al$. The reported concomitant maximum in $A$ value is also observable from the table, indicating that the nanoparticles also behave much in the same way as the bulk in terms of quantum criticality. However, the minimum value of $n$ does not enter the non-Fermi liquid (NFL) range (1.56 $\textless$ n $\textless$ 2), and hence it seems that the material does not possess a quantum critical state. We can not make any decisive statement about the quantum criticality of the nanoalloys here, because we neither have sufficient number of $x$ values near QC nor enough $\rho$-T data points in the low-T region to get more accurate fits. As a further check, therefore, we performed the DC magnetization measurements on the alloys near $x_c$, to be discussed in the following section.

\begin{table}
\caption{\label{tab1}Values of the fitting parameters $A$ and $n$ in the Pd$_{1-x}$Ni$_x$ nanoalloys.}
\centering
\begin{tabular}{lll}
\br
$x$&$A (n\Omega /K)$&$n$\\
\br
0.02&0.11$\pm$ 0.01&2.2$\pm$ 0.04\\
0.026&0.60$\pm$ 0.10&2.1$\pm$ 0.06\\
0.03&0.21$\pm$ 0.02&2.2$\pm$ 0.04\\
0.16&0.02$\pm$ 0.002&2.9$\pm$ 0.04\\
\br
\end{tabular}\\

\end{table}

\subsection{Magnetization}

\begin{figure}%
\centering
\subfigure[][]{%
\label{fig:7-a}%
\includegraphics[width=0.4\textwidth]{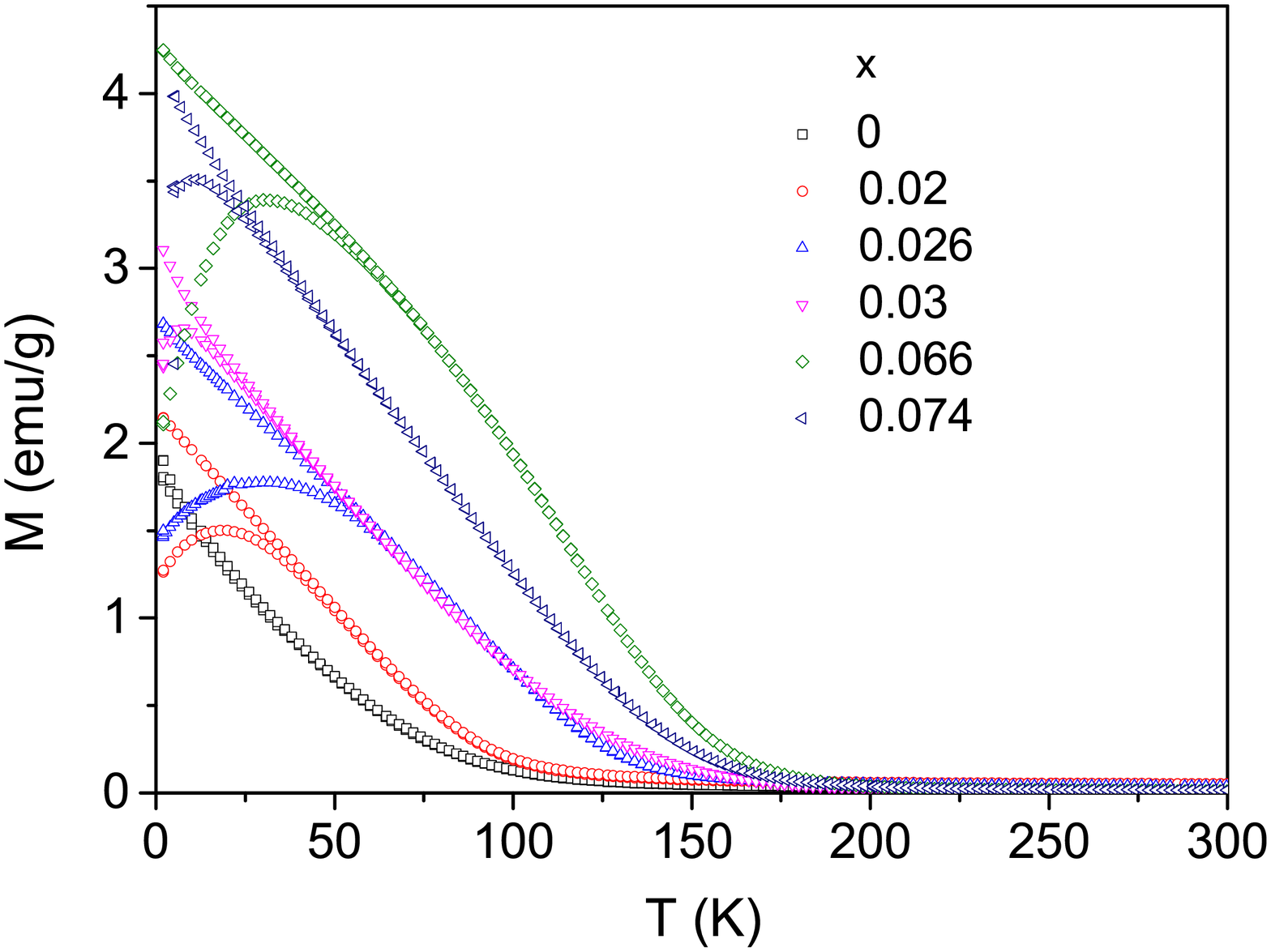}}%
\hspace{8pt}%
\subfigure[][]{%
\label{fig:7-b}%
\includegraphics[width=0.4\textwidth]{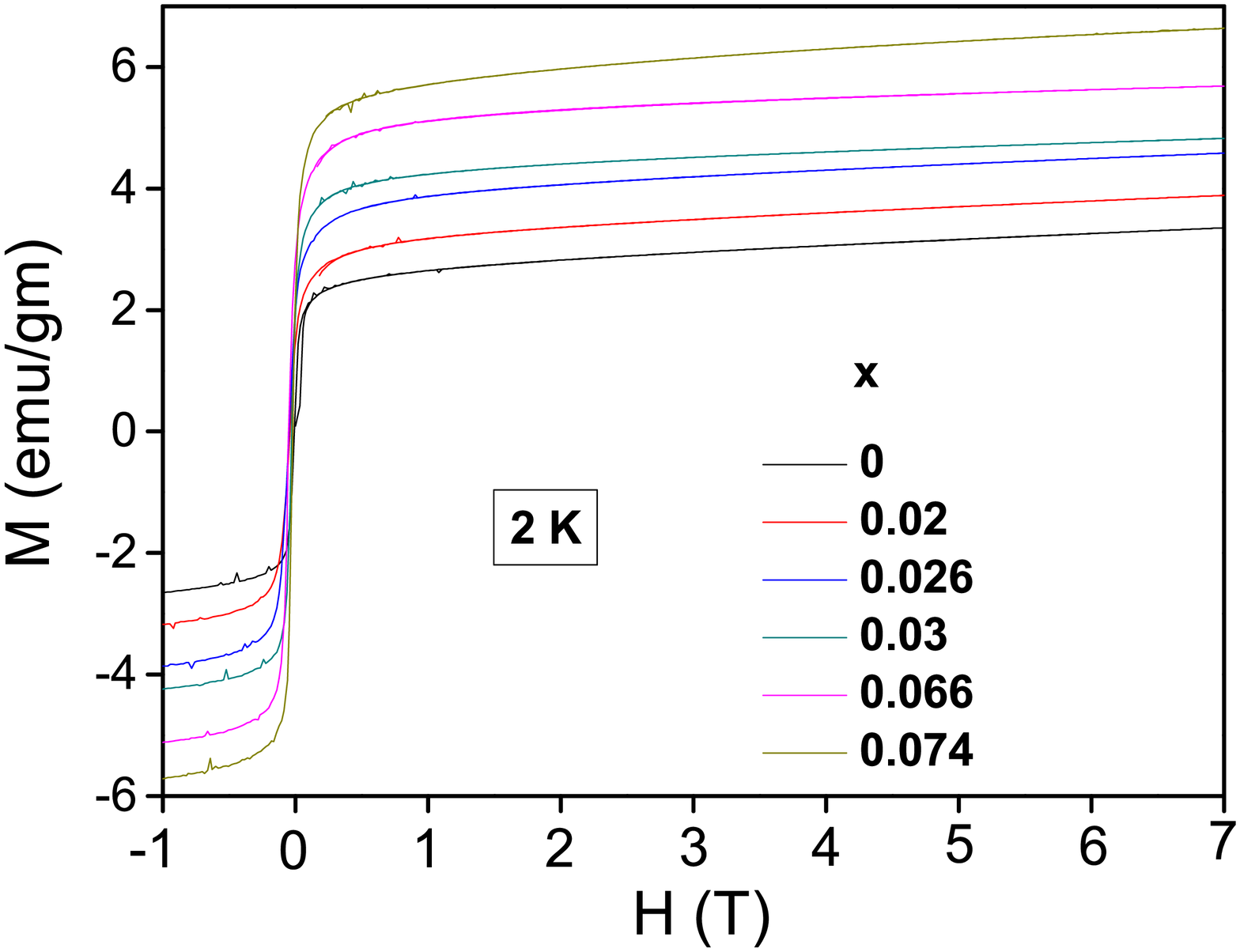}}%
\hspace{8pt}%
\subfigure[][]{%
\label{fig:7-c}%
\includegraphics[width=0.4\textwidth]{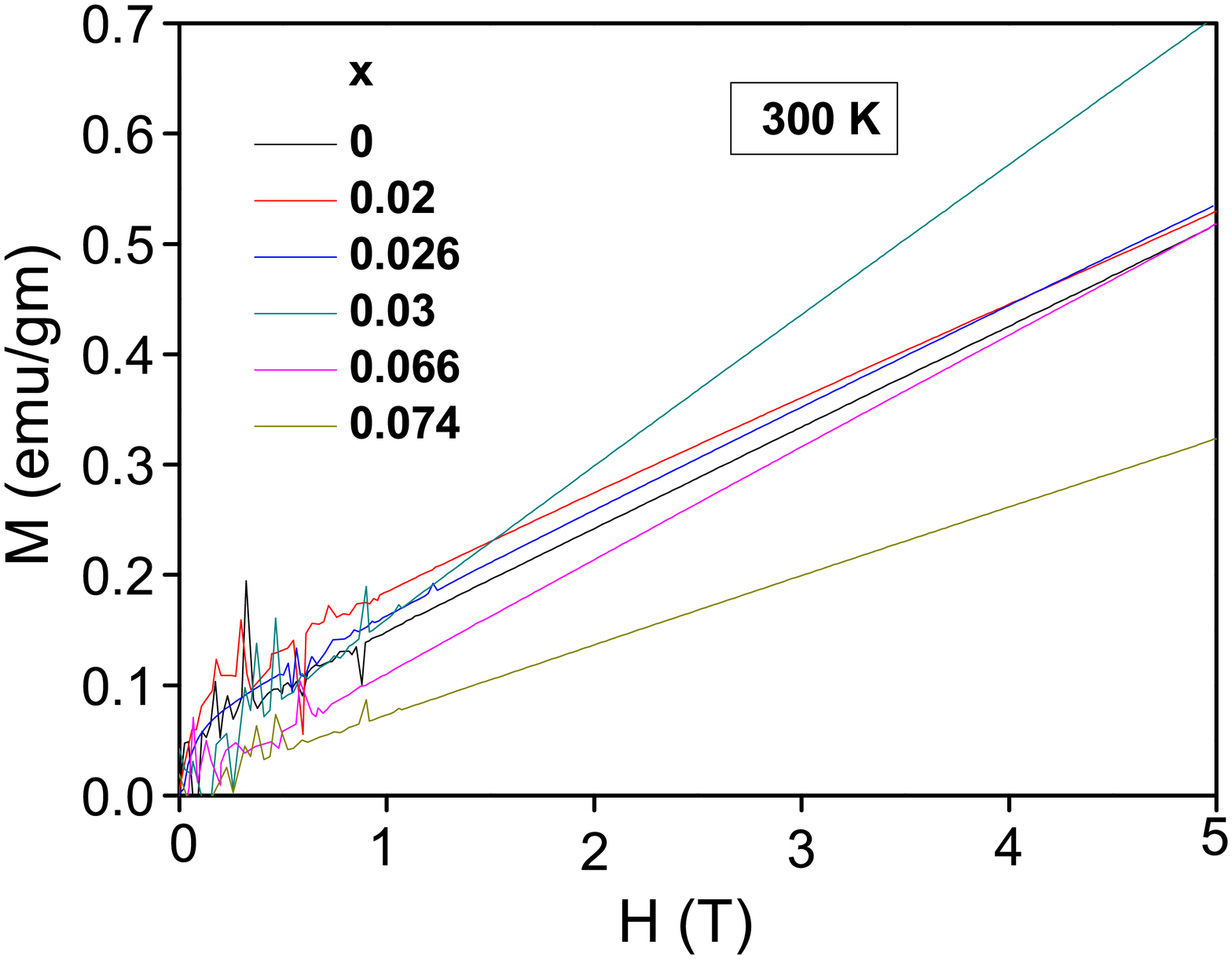}}%

\caption[]{
\subref{fig:7-a} FC and ZFC magnetizations versus temperature at 500 Oe filed.
\subref{fig:7-b} M-H curves at 2 K.
\subref{fig:7-c} M-H curves at 300 K.}
\label{fig:7}%
\end{figure}

The temperature dependences of the field cooled (FC) and zero-field cooled (ZFC) DC magnetizations in 2 K $\textless~T~\textless$ 300 K temperature range and at 500 Oe magnetic field are shown in Fig. \ref{fig:7-a} for compositions 0 ${\leq} ~x ~{\leq}$ 0.076. Each curve exhibits a FC-ZFC splitting with a maximum, except for pure Pd, in the ZFC curve at a temperature $T_{max}$ and an irreversibility point $T_{irr} ~\textgreater ~T_{max}$; the curve for pure Pd does not show $T_{max}$ down to 2 K. The FC curves for all the samples, however, increase monotonically with decreasing temperature. These observations indicate the presence of magnetic nanoparticles with size and shape dispersions, as also evidenced by the HRTEM images, in all the samples \cite{Arne}. Broadly, the system is in a blocked ferromagnetic (FM) state below $T_{irr}$, is in a superparamagnetic (SPM) state up to the temperature $T_C$ beyond which the M-T curve starts obeying the Curie-Weiss law, and then even the individual SPM particles become paramagnetic (PM) \cite{Arne, Castrillon}. $T_C$'s in the present study have been determined simply by finding the lowest temperature to which the high-T part of the M-T curve fits well with a Curie-Weiss law. The FC or ZFC curves for all the samples, however, do not show a distinct inverse temperature dependence between $T_{irr}$ and $T_C$; they quite apparently have a power-law temperature dependence component, akin to ferromagnetism, as well. The samples are thus in a mixed FM and SPM state in the range $T_{irr}~\textless~T~\textless~T_C$. The presence of FM particles at low temperatures is evidenced by the appearance of finite remanance and coercivity in the field dependence of magnetization (M-H curve) of all the samples at 2 K (Fig. \ref{fig:7-b}), while their transition to PM particles is revealed by the absence of the remanence and coercivity along with the non-saturation of magnetization at 300 K (Fig. \ref{fig:7-c}) in all the cases. Since the magnetization does not saturate with field in the whole measured temperature range, the presence of PM particles at all temperatures can also not be denied.

\begin{figure}%
\centering
\subfigure[][]{%
\label{fig:8-a}%
\includegraphics[width=0.4\textwidth]{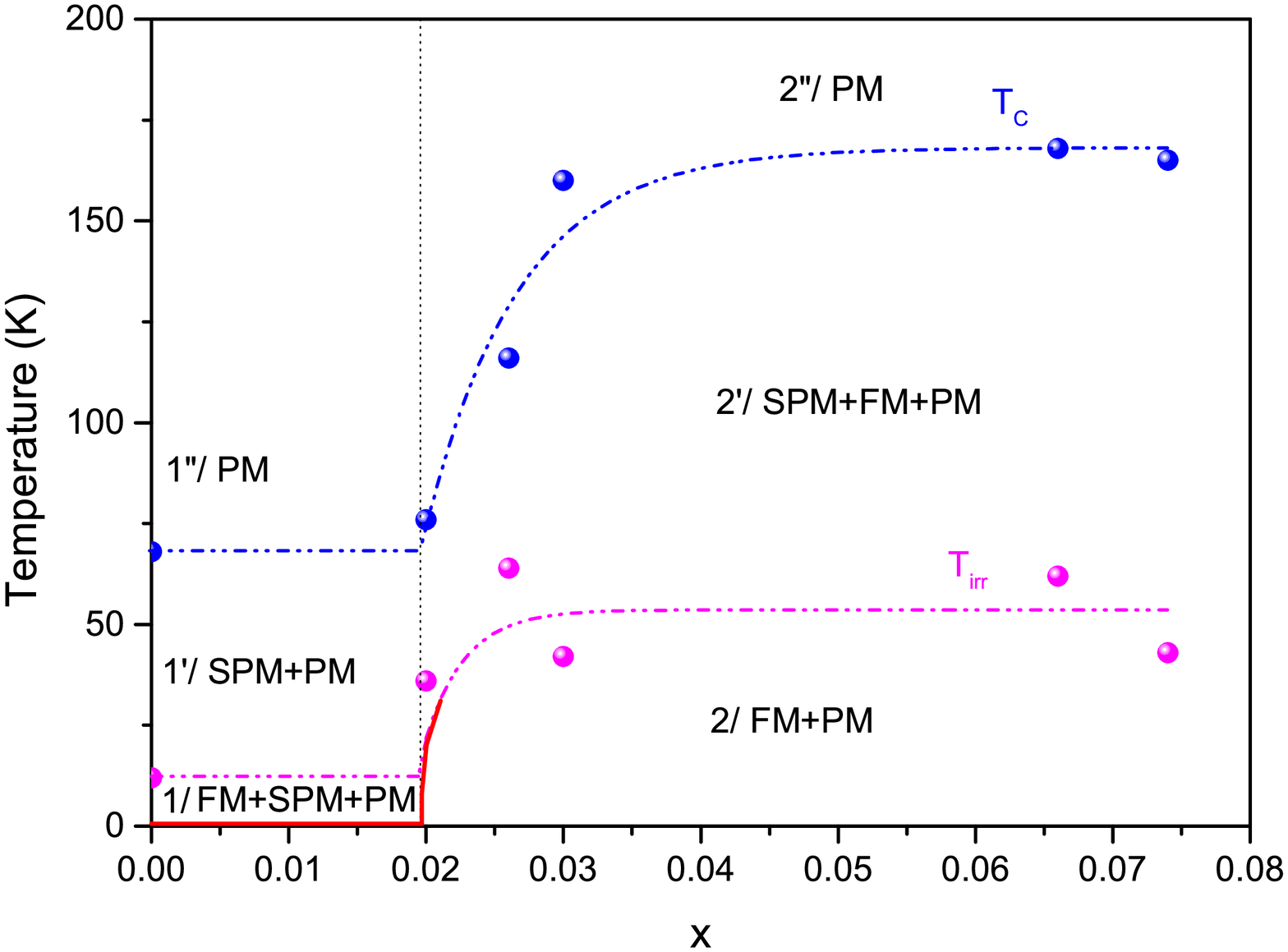}}%
\hspace{8pt}%
\subfigure[][]{%
\label{fig:8-b}%
\includegraphics[width=0.45\textwidth]{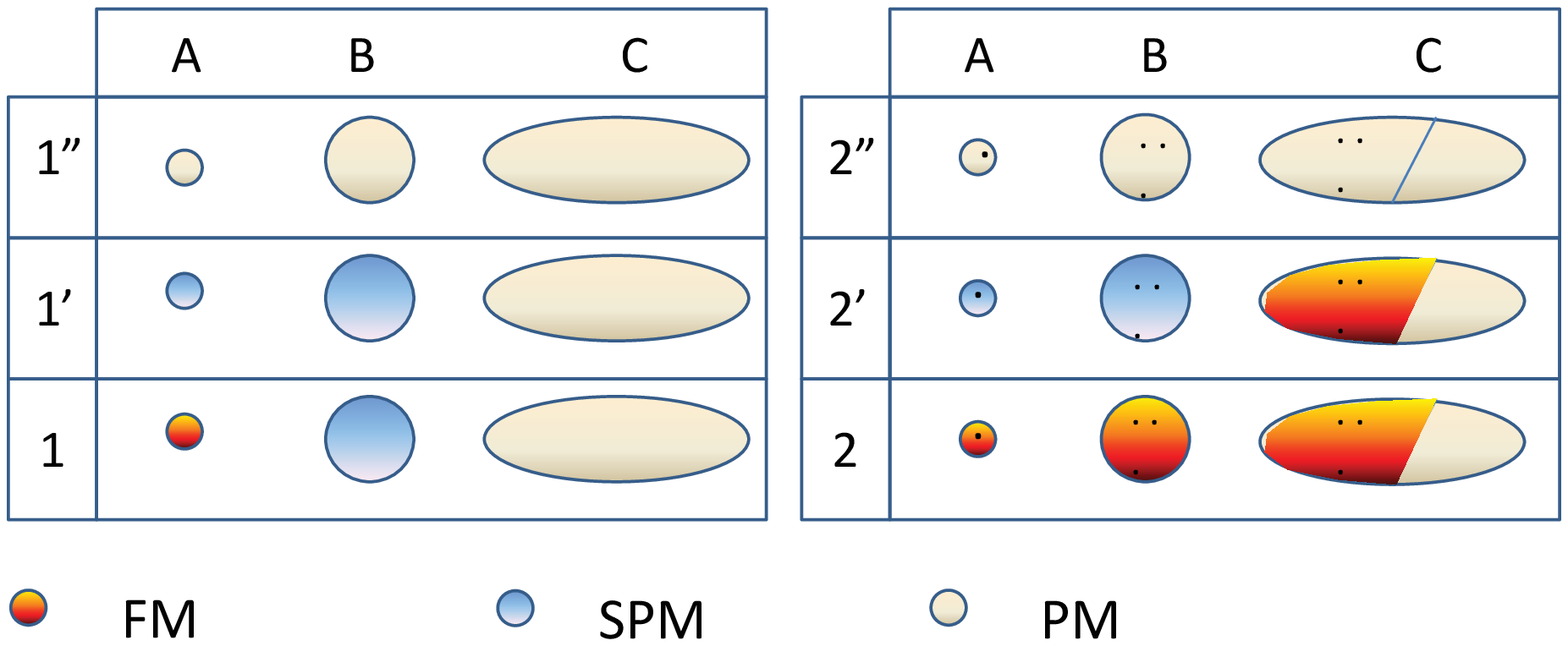}}%

\caption[]{
\subref{fig:8-a} T-x phase diagram. The thick red curve indicates the anticipated QPT along T = 0 axis.
\subref{fig:8-b} Schematic of the particles present in the system and their magnetic states.}
\label{fig:8}%
\end{figure}

A tentative $T-x$ phase diagram is drawn from the above analyses, as shown in Fig. \ref{fig:8-a}. There exist two phase boundaries - one determined by $T_C$  and the other by $T_{irr}$ - along the temperature axis. The $T_C$ and $T_{irr}$ lines can be fitted above $x$ = 0.02, the minimum Ni concentration taken in the present study, by an $a-bc^x$ dependence, where $a$, $b$, and $c$ are some constants not relevant to be shown here. Below $x$ = 0.02, the two curves are quite apparently flat horizontal lines. Thus, the whole phase diagram can be divided into six regions - 1, 1', 1", 2, 2', and 2", as shown in the figure. Based on the arguments above, the regions 2, 2' and 2" can be ascribed to FM+PM, SPM+FM+PM and PM phases, respectively. The magnetic behaviour in the regions 1, 1' and 1" can be understood by the behaviour of pure Pd as a representative of compositions $x$ = 0 to 0.02. For this, we propose the following: we divide the nanoparticles into three size ranges - small ($\textless$ 20 nm), 20-50 nm, and elongated ones, represented in the schematic diagram, Fig. \ref{fig:8-b}, as particles A, B and C, respectively. The particle A in region 1 is ferromagnetic, as suggested by Refs. \cite{Angappane08, Coronado08}. The particle B in region 1 is proposed to be of superparamagnetic or paramagnetic kind.

The particle C, however, is elongated and hence must behave like bulk Pd, i.e., it must be paramagnetic in nature. In the region 1', A transforms to a mixed SPM+PM state, as evidenced by the corresponding M-T curve, while B and C remain SPM/PM. In region 1", however, A, B, C all become paramagnetic. In the region 2, since all the particles contain Ni atoms now, A and B become ferromagnetic. For the C particles, we may assume the presence of some regions a bit richer in Ni content than the corresponding $x$, while others having no Ni atoms, since Ni can occupy random positions in the alloy. Such statistical clusterings of Ni atoms in Pd$_{1-x}$Ni$_x$ alloys has also been shown in a recent study of a muon spin relaxation study of magnetism in Pd$_{1-x}$Ni$_x$ bulk alloys near $x_c$ by Kalvius {\it et al.} \cite{Kalvius 2014}. The Ni-rich regions are then ferromagnetic while the Ni-less regions are paramagnetic, as shown in the Fig. \ref{fig:8-b}. Then beyond $T_{irr}$ and in the region 2', the A particles remain FM, the B particles become superparamagnetic, and the C particles remain FM+PM in nature. In region 2", then, all the particles become paramagnetic. We propose that if we can prepare just the particles with intermediate sizes B, then the horizontal $T_{irr}$ line for $x~\textless$ 0.02 in the Fig. \ref{fig:8-a} can be pulled down to 0 K in the absence of the FM smaller particles. In this situation, a SPM to PM phase transition can be attained at 0 K at a critical concentration around $x$ = 0.02. The minimum value of the exponent $n$ in the resistivity fitting, which is 2 in the present case, must be due to the predominance of electron-magnon $T^2$ term \cite{White59} over the NFL's $\textless$ 2 value because of the presence of the FM A particles. We anticipate that this value can be brought down to $\textless$ 2 by ruling out the electron-magnon scattering in the absence of A particles. This anticipation of attaining QC in the Pd$_{1-x}$Ni$_x$ nanoparticles is, however, more speculative than definitive, and some experiments with monodispersed intermediate-sized Pd$_{1-x}$Ni$_x$ nanoparticles can be done to verify the existence of QC in these nanoalloys.

\section{Conclusions}

Pd$_{1-x}$Ni$_x$ (0.01 $\leq x \leq$ 0.50) nanoalloys were synthesized by a chemical reflux method using diethanolamine as the surfactant and hydrazine hydrate as the reducing agent. The finally prepared compositions were determined by EDAX and were found to scale linearly with the initial, aimed compositions. The FESEM, HRTEM, XRD and XPS results showed that the particles for all the samples were of 40-50 nm mean diameter, were crystalline and had pure Pd-Ni alloy phases without any trace of unreacted Pd or Ni or any oxide. Further, the high-T part of the temperature dependence of resistivity confirmed the metallic nature of all the samples. In addition, a fit of the low- resistivity with $AT^n$ showed an $x$-dependent upturn in the $n$ value at $x$ = 0.026, the bulk QCC, with a minimum value of 2.2, a value attributable to Fermi liquids. There was, however, a concomitant upturn in the $A$ value, indicating the presence of a QPT-like behaviour in the material. The DC magnetization results suggested a tentative T-$x$ phase diagram separated into three regions by the paramagnetic Curie temperature $T_C$ and the temperature $T_{irr}$ of irreversibility between the FC and ZFC magnetizations. Each of the three regions is further subdivided into 0 $\leq x \leq$ 0.02 and $x\textgreater$ 0.02 regions. The magnetic behaviours of the system in each of these six sub-regions is explained by a proposed subdivision of the particles into small, medium and elongated nanoparticles. It is further anticipated that with a sample with monodispersed medium-sized nanoparticles, there exists a possibility to observe a QPT in these nanoalloys.

\vspace{10 mm}

{\bf ACKNOWLEDGEMENTS}
\vspace{10 mm}

We sincerely acknowledge A. Banerjee and R. Rawat of UGC-DAE Consortium for Scientific Research, Indore for the magnetization and resistivity measurements, respectively. P. Swain would also like to acknowledge the financial support from CSIR, New Delhi.

\end{document}